\documentclass[prd,amsmath,superscriptaddress,nofootinbib,showpacs,preprintnumbers]{revtex4}
\usepackage{bm}
\usepackage{amsmath}
\usepackage[dvips]{graphicx}

\begin{document}

\title{Chiral crossover, deconfinement and quarkyonic matter within a Nambu-Jona Lasinio model with the Polyakov loop}
\author{H. Abuki}\email{hiroaki.abuki@ba.infn.it}
\affiliation{I.N.F.N., Sezione di Bari, I-70126 Bari, Italia} \affiliation{Universit\`a di Bari, I-70126 Bari, Italia}
\author{R. Anglani}\email{roberto.anglani@ba.infn.it}
\affiliation{I.N.F.N., Sezione di Bari, I-70126 Bari, Italia} \affiliation{Universit\`a di Bari, I-70126 Bari, Italia}
\author{R.Gatto}\email{raoul.gatto@physics.unige.ch}
\affiliation{D\'epartement de Physique Th\'eorique, Universit\'e de Gen\`eve, CH-1211 Gen\`eve 4, Switzerland}
\author{G. Nardulli}\email{giuseppe.nardulli@ba.infn.it}
\affiliation{I.N.F.N., Sezione di Bari, I-70126 Bari, Italia} \affiliation{Universit\`a di Bari, I-70126 Bari, Italia}
\author{M. Ruggieri}\email{marco.ruggieri@ba.infn.it}
\affiliation{I.N.F.N., Sezione di Bari, I-70126 Bari, Italia} \affiliation{Universit\`a di Bari, I-70126 Bari, Italia}

\preprint{BA-TH/591-08} 

\begin{abstract}
We study the interplay between the chiral and the deconfinement transitions, both at high temperature and high quark
chemical potential, by a non local Nambu-Jona Lasinio model with the Polyakov loop in the mean field approximation and
requiring neutrality of the ground state. We consider three forms of the effective potential of the Polyakov loop: two
of them with a fixed deconfinement scale, cases I and II, and the third one with a $\mu$ dependent scale, case III. In
the cases I and II, at high chemical potential $\mu$ and low temperature $T$ the main contribution to the free energy
is due to the $Z(3)$-neutral three-quark states, mimicking the quarkyonic phase of the large $N_c$ phase diagram. On
the other hand in the case III the quarkyonic window is shrunk to a small region. Finally we comment on the relations
of these results to lattice studies and on possible common prospects. We also briefly comment on the coexistence of
quarkyonic and color superconductive phases.
\end{abstract}

\maketitle

\section{Introduction}
Color confinement and chiral symmetry breaking are some of the most intriguing topics in modern theoretical physics.
Quantum Chromodynamics (QCD) is believed to be the ultimate theory describing strong interactions. Nowadays it is
accepted that the main ground state properties of QCD can be described in terms of non perturbative spontaneous
breaking and/or restoring of some of the global symmetries of the QCD lagrangian.

Unfortunately solving QCD in its non perturbative regime is a hard task. At zero and small quark chemical potential
$\mu$ lattice calculations are a good tool to derive the equation of state of QCD matter, the transition temperatures
and so on starting from the first principles, see for
example~\cite{Aoki:2006br,Schmidt:2006us,Philipsen:2005mj,Heller:2006ub} and references therein. Several approximation
methods are available to overcome the sign problem of the fermion determinant with three colors at finite $\mu$ (see
Refs.~\cite{Ejiri:2004yw,Splittorff:2006vj,Splittorff:2006fu} for reviews on the sign problem): small-$\mu$
expansion~\cite{Allton:2003vx,Allton:2002zi,Allton:2005gk}, reweighting tecniques~\cite{Fodor:2001pe,Fodor:2002km},
density of the states methods~\cite{Fodor:2007vv} and analytic continuation to imaginary chemical
potential~\cite{Laermann:2003cv,de Forcrand:2003hx,D'Elia:2007ke,D'Elia:2004at,D'Elia:2002gd}.

Besides lattice calculations there exist effective descriptions of QCD. Among them Nambu-Jona Lasinio (NJL in the
following) models~\cite{Nambu:1961tp} are very popular, see~\cite{revNJL} for reviews. They are based on the
observation that several properties of the QCD ground state are related to the spontaneous breaking of some of the
global symmetries of the QCD lagrangian. Therefore one hopes that by a model that has the same global symmetry breaking
of QCD one can capture the essential physics of QCD itself.

In recent years it has been argued that the NJL model, which does not contain gluons, can be improved by adding a non
linear term to the lagrangian which describes the dynamics of the traced Polyakov loop~\cite{Polyakovetal}, and an
interaction term of the Polyakov loop with the quarks. The resulting model is called the PNJL model, introduced in
Refs.~\cite{Meisinger:1995ih,Fukushima:2003fw} and extensively studied
in~\cite{Ratti:2005jh,Roessner:2006xn,Ghosh:2007wy,Kashiwa:2007hw,
Schaefer:2007pw,Ratti:2007jf,Sasaki:2006ww,Megias:2006bn,Zhang:2006gu,Fukushima:2008wg,Sakai:2008py,Sakai:2008um,Kashiwa:2008ga,
Ciminale:2007ei,Fu:2007xc,Ciminale:2007sr,Hansen:2006ee,Abuki:2008tx,Abuki:2008ht,Contrera:2007wu,Blaschke:2007np}. In
the PNJL model one assumes that a homogeneous euclidean temporal background gluon field couples to the quarks via the
covariant derivative of QCD. This coupling gives rise to the interplay between the chiral condensate and the Polyakov
loop. Even if it is very simple, the PNJL model turned out to be a powerful tool which allows to compute several
quantities that can be computed on the lattice as well. The agreement with existing lattice data is
satisfactory~\cite{Ratti:2005jh}.

One of the exciting characteristics of the PNJL model is the {\em statistical confinement} of quarks at low
temperature~\footnote{The term statistical confinement has been introduced by K. Redlich during the Workshop ``New
Frontiers in QCD08''.}. In a few words this means that at small temperature and small chemical potential the
contribution to the free energy, $\Omega$, of the states with one and two quarks are suppressed, and the leading
contribution to $\Omega$ arises from the thermal excitations of colorless three quark states. This property is related
to the small value of the expectation value of Polyakov loop which is found in the self-consistent calculations within
the PNJL model in the aforementioned conditions of temperature and chemical potential. It has been recently argued by
Fukushima that the statistical confinement property of the PNJL model persists even at high chemical
potential~\cite{Fukushima:2008wg}. This result is in agreement with the phase diagram of QCD obtained in the large
number of colors $N_c$ approximation~\cite{McLerran:2007qj,Hidaka:2008yy}, see also
Refs.~\cite{Glozman:2007tv,Glozman:2008kn} for recent related studies. Inspired by Ref.~\cite{McLerran:2007qj}
Fukushima has suggested to interpret the statistical confined phase of the PNJL model at high quark chemical potential
as the quarkyonic state found in~\cite{McLerran:2007qj}.

In this work we investigate on the ground state of the electrically neutral two flavor PNJL model, focusing on its
possible quarkyonic structure at high $\mu$ and low $T$. We use a non local four fermion interaction instead of the
local one~\cite{Nambu:1961tp}. The local NJL model is usually regularized by means of an ultraviolet sharp cutoff,
which amounts to artificially cutoff the quark momenta that are larger than the cutoff itself. Thus the extensions of
the model to temperatures and/or chemical potentials of the order of the cutoff are quite dubious. However if one
introduces a non local interaction, which corresponds to the multiplication of the NJL coupling by a momentum dependent
form factor $f(p)$, and requires that the form factor satisfies the asymptotic freedom property of QCD
$f(p\rightarrow\infty)=0$, then all of the momentum integrals are convergent and the model is consistent at any value
of temperature and chemical potential. In this paper we use one specific form of the form factor. Although the choice
of a different functional form for $f(p)$ can lead to different quantitative results (mainly the shift of the critical
points) we believe that our picture should not be modified qualitatively. We consider the logarithmic form of the
Polyakov loop effective potential ${\cal U}$ suggested by Ratti, Roessner and Weise in Ref.~\cite{Roessner:2006xn};
moreover we investigate on the effects of a dependence of ${\cal U}$ on the quark chemical potential as well as on the
number of flavors as suggested in Ref.~\cite{Schaefer:2007pw}. We compare the phase diagrams obtained in the cases in
which we do not consider (cases I and II) and do consider (case III) the $\mu$-dependence of ${\cal U}$. Cases I and II
differ for the value of the deconfinement scale in the Polyakov loop effective potential.

We find that the phase diagram in the two cases (I and II on one side, III on the other side) differ even
qualitatively. In particular, on one hand in the cases I and II we confirm the results of
Fukushima~\cite{Fukushima:2008wg} and strengthen his interpretation of the high chemical potential/small temperature
state of the PNJL model as the quarkyonic matter of the large $N_c$ phase diagram. On the other hand in case III we
find that the quarkyonic-like window found in the cases I and II is shrunk and becomes a small region in the $\mu-T$
plane in the case III, opening a wide room for the deconfined quark matter of the pure NJL model.

The plan of the paper is as follows. In Section II we sketch the formalism. In Section III we discuss our results.
Finally in Section IV we draw our conclusions.

\section{Thermodynamic potential with a non local four fermion interaction}
The Lagrangian density of the two flavor PNJL model is given by~\cite{Fukushima:2003fw,Abuki:2008tx}
\begin{equation}
{\cal L}^\prime= \bar{e}(i\gamma_\mu\partial^\mu)e +  \bar\psi\left(i\gamma_\mu D^\mu + \mu\gamma_0 -m\right)\psi +
{\cal L}_{4} - {\cal U}[\Phi,\bar\Phi,T]~. \label{eq:LagrP}
\end{equation}
In the above equation $e$ denotes the electron field;  $\psi$ is the quark spinor with Dirac, color and flavor indices
(implicitly summed). $m$ corresponds to the bare quark mass matrix; we assume from the very beginning $m_u = m_d$. The
covariant derivative is defined as usual as $D_\mu =
\partial_\mu -i A_\mu$. The gluon background field
$A_\mu=\delta_{0\mu}A_0$ is supposed to be homogeneous and static, with $A_0 = g A_0^a T_a$ and $T_a$, $a=1,\dots,8$
being the $SU(3)$ color generators with the normalization condition $\text{Tr}[T_a,T_b]=\delta_{ab}$.  Finally $\mu$ is
the chemical mean quark chemical potential, related to the conserved baryon number.

\begin{widetext}
In Eq.~\eqref{eq:LagrP} $\Phi$, $\bar\Phi$ correspond to the normalized traced Polyakov loop $L$ and its hermitian
conjugate respectively, $\Phi=\text{Tr}W/N_c$, $\bar\Phi=\text{Tr}W^\dagger/N_c$, with
\begin{equation}
W={\cal P}\exp\left(i\int_0^\beta A_4 d\tau\right)=\exp\left(i \beta A_4\right)~,~~~~~A_4=iA_0~,
\end{equation}
and $\beta=1/T$. $\Phi$ is a color singlet but it has a $Z(3)$ charge~\cite{Polyakovetal}, where $Z(3)$ is the center
of the color group $SU(3)$; thus if $\Phi\neq0$ in the ground state then the $Z(3)$ symmetry is spontaneously broken.
The term ${\cal U}[\Phi,\bar\Phi,T]$ is the effective potential for the traced Polyakov loop; in absence of dynamical
quarks it is built in order to reproduce the pure glue lattice data of QCD, namely thermodynamical quantities
(pressure, entropy and energy density) and the deconfinement temperature of heavy (non-dynamical) quarks, $T= 270$ MeV.
Several forms of this potential have been suggested in the literature, see for
example~\cite{Fukushima:2003fw,Ratti:2005jh,Roessner:2006xn,Ghosh:2007wy,Fukushima:2008wg}. In this paper we adopt the
following logarithmic form~\cite{Roessner:2006xn},
\begin{equation}
{\cal U}[\Phi,\bar\Phi,T] = T^4\left[-\frac{b_2(T)}{2}\bar\Phi\Phi + b(T)\log\left[1-6\bar\Phi\Phi + 4(\bar\Phi^3 +
\Phi^3) -3(\bar\Phi\Phi)^2\right]\right]~,\label{eq:Poly}
\end{equation}
with
\begin{equation}
b_2(T) = a_0 + a_1 \left(\frac{\bar T_0}{T}\right) + a_2 \left(\frac{\bar T_0}{T}\right)^2~,~~~~~b(T) =
b_3\left(\frac{\bar T_0}{T}\right)^3~.\label{eq:lp}
\end{equation}
Numerical values of the coefficients are as follows~\cite{Roessner:2006xn}:
\begin{equation}
a_0=3.51~,~~~a_1 = -2.47~,~~~a_2 = 15.2~,~~~b_3=-1.75~.
\end{equation}
If dynamical quarks were not present then one should chose $\bar T_0 = 270$ MeV in order to reproduce the deconfinement
transition at $T = 270$ of the pure gauge theory~\cite{Fukushima:2003fw,Ratti:2005jh,Ratti:2007jf}.  In presence of
quarks $\bar T_0$ might get a dependence on the number of active flavors as well as on the quark chemical
potential~\cite{Ratti:2005jh,Schaefer:2007pw}. Inspired by Refs.~\cite{Fukushima:2003fw,Ratti:2005jh,Schaefer:2007pw}
in this paper we consider three cases:
\begin{eqnarray}
\bar{T}_0 &=& 208~\text{MeV}~,~~~\text{Case I}~,\\
\bar{T}_0 &=& 270~\text{MeV}~,~~~\text{Case II}~,\\
\bar{T}_0(\mu) &=& T_\tau e^{-1/\alpha_0 c(\mu)}~,~~~\text{Case III}~.\label{eq:T0m}
\end{eqnarray}
Case II corresponds to the deconfinement temperature in the pure glue theory; the parameters in the cases I and III
have been evaluated in Ref.~\cite{Schaefer:2007pw} on the basis of hard dense and hard thermal loop approximations to
QCD. In the equation corresponding to Case III we have set
\begin{equation}
\alpha_0 = 0.304~,~~~T_\tau = 1770~\text{MeV}~,
\end{equation}
and
\begin{equation}
c(\mu) = \frac{11 N_c - 2 N_f}{6\pi} - \frac{16 N_f}{\pi}\frac{\mu^2}{T_\tau^2}~,
\end{equation}
\end{widetext}
with $N_f = 2$ and $N_c = 3$. At $\mu=0$ we have $\bar{T}_0(\mu=0) = 208$ MeV as case I; for comparison, at $\mu=500$
MeV the deconfinement scale is given by $\bar{T}_0(\mu=0) = 19$ MeV.

In Eq.~\eqref{eq:LagrP} ${\cal L}_4$ represents the lagrangian density for the four fermion interaction. If we define
$S_4 = \int d^4 x {\cal L}_4$ as the interaction action then in the local version of the NJL model one has
\begin{equation}
S_{4} = G \int d^4 x~\left[(\bar\psi \psi)^2 + (\bar\psi i\gamma_5\bm\tau\psi)^2\right]~. \label{eq:c1}
\end{equation}
In the non local version of the NJL model the contact term Eq.~\eqref{eq:c1} is replaced
by~\cite{Sasaki:2006ww,Schmidt:1994di,Bowler:1994ir,Blaschke:2000gd,GomezDumm:2005hy,Aguilera:2006cj,Grigorian:2006qe}
\begin{equation}
S_{4} = G \int d^4 x~\left[(\bar q(x) q(x))^2 + (\bar q(x) i\gamma_5\bm\tau q(x))^2\right]~,\label{eq:1}
\end{equation}
where the dressed quark field is defined as
\begin{equation}
q(x)= \int d^4 y~F(x-y) \psi(y)~,\label{eq:dress}
\end{equation}
and $F(r)$ is a form factor whose Fourier transform $f(p)$ satisfies the constraint $f(p)\rightarrow 0$ for
$p\rightarrow\infty$, $p$ being the 3-momentum. In this paper we follow Ref.~\cite{Sasaki:2006ww} and use the
Lorentzian form factor,
\begin{equation}
f(p)=\frac{1}{\sqrt{1+(p/\Lambda)^{2\alpha}}}~.\label{eq:f1}
\end{equation}
In the above equation $\Lambda = 684.2$ MeV and $\alpha=10$. Moreover we use $m=4.46$ MeV and
$G=2.33/\Lambda^2$~\cite{Sasaki:2006ww}. By these numerical values we reproduce the pion decay constant $f_\pi = 92.3$
MeV and the pion mass $m_\pi = 135$ MeV, as well as the chiral condensate $\langle\bar u u\rangle =
-(256.2~\text{MeV})^3$. Although the choice of a different form factor will lead to different critical temperatures
and/or chemical potentials, it is quite reasonable that the qualitative picture that we draw in this work is
insensitive to the specific form of $f(p)$.

As explained in the Introduction we are interested to the ground state of the model specified by the Lagrangian in
Eq.~\eqref{eq:LagrP}, at each value of the temperature $T$ and the chemical potential $\mu$,  corresponding to a
vanishing total electric charge. In order to build the neutral ground state we use the standard grand canonical
ensemble formalism, adding to Eq.~\eqref{eq:LagrP} the term $\mu_Q N_Q$, $\mu_Q$ being the chemical potential (i.e.
Lagrange multiplier) for the total charge $N_Q$, and requiring stationarity of the thermodynamic potential with respect
to variations of $\mu_Q$, which is equivalent to the requirement $<N_Q>=0$ in the ground state. This amounts to write
the lagrangian ${\cal L}$ in the gran canonical ensemble ${\cal L} = {\cal L}^\prime + \mu_Q N_Q$
as~\cite{Abuki:2008tx}
\begin{equation}
{\cal L}=\bar{e}(i\gamma_\mu\partial^\mu + \mu_e \gamma_0)e +  \bar\psi\left(i\gamma_\mu D^\mu + \hat\mu\gamma_0
-m\right)\psi + G\left[\left(\bar\psi \psi\right)^2 + \left(\bar\psi i \gamma_5 \vec\tau \psi\right)^2\right] - {\cal
U}[\Phi,\bar\Phi,T]~, \label{eq:Lagr}
\end{equation}
where $\mu_e = - \mu_Q$ and the quark chemical potential matrix $\hat\mu$ is defined in flavor-color space as
\begin{equation}
\hat\mu=\left(\begin{array}{cc}
  \mu-\frac{2}{3}\mu_e & 0 \\
  0 & \mu + \frac{1}{3}\mu_e \\
\end{array}\right)\otimes\bm{1}_c~,\label{eq:chemPot}
\end{equation}
where $\bm{1}_c$ denotes identity matrix in color space. At $\mu_e\neq0$ a difference of chemical potential between up
and down quarks, $\delta\mu=\mu_2 /2$, arises.

In this paper we work in the mean field approximation. Because of $\delta\mu\neq 0$ a pion condensation might occur in
the ground state~\cite{Ebert:2005wr}. In order to study simultaneously chiral symmetry breaking and pion condensation
we assume that in the ground state the expectation values, real and independent on $x$, for the following operators may
develop~\cite{Zhang:2006gu,Abuki:2008tx,Ebert:2005wr,Ebert:2000pb,Ebert:2008tp},
\begin{equation}
\sigma = G\langle \bar q(x)q(x)\rangle~,~~~ \pi = G\langle \bar q(x)i\gamma_5\tau_1 q(x)\rangle~.\label{eq:condensates}
\end{equation}
In the above equation a summation over flavor and color is understood.  We have assumed that the pion condensate aligns
along the $\tau_1$ direction in flavor space. This choice is not restrictive. As a matter of fact we should allow for
independent condensation both in $\pi^+$ and in $\pi^-$ channels~\cite{Zhang:2006gu}:
\begin{equation}
\pi^\pm\equiv G\langle\bar\psi i \gamma_5 \tau_\pm \psi\rangle = \frac{\pi}{\sqrt{2}}e^{\pm i\theta}~,
\end{equation}
with $\tau_\pm = (\tau_1\pm\tau_2)/\sqrt{2}$; but the thermodynamical potential is not dependent on the phase $\theta$,
therefore we can assume $\theta=0$ which leaves us with $\pi^+ = \pi^- = \pi/\sqrt{2}$ and introduce only one
condensate, specified in Eq.~\eqref{eq:condensates}.

In what follows we consider the system at finite temperature $T$ in the volume $V$. This implies that the space-time
integral is $\int d^4x = \int_{0}^\beta d\tau \int d^3\bm x$ with $\beta=1/T$. In the mean field approximation the PNJL
action reads
\begin{eqnarray}
S&=&\int d^4 x\left[\bar{e}(i\gamma_\mu\partial^\mu + \mu_e \gamma_0)e + \bar\psi\left(i\gamma_\mu D^\mu +
\hat\mu\gamma_0 \right)\psi\right] \nonumber\\
&&+2 \sigma\int d^4 x~\bar q(x) q(x) + 2\pi\int d^4 x~\bar q(x)i\gamma_5 \tau_1 q(x)\nonumber\\
&&
 ~~~- \beta V \frac{\sigma^2 + \pi^2}{G} - \beta V {\cal U}[\Phi,\bar\Phi,T]~, \label{eq:LagrMF}
\end{eqnarray}
where $V$ is the quantization volume and $\beta=1/T$. In momentum space one has
\begin{eqnarray}
S&=&\int\frac{d^4p}{(2\pi)^4} \left[\bar{e}(\gamma_\mu p^\mu + \mu_e \gamma_0)e + \bar\psi\left(\gamma_\mu p^\mu
-\gamma_\mu A^\mu - \hat\mu\gamma_0 \right)\psi\right] \nonumber\\
&&+\int\frac{d^4p}{(2\pi)^4} f(p)^2\left[2\sigma~\bar\psi(p) \psi(p) + 2\pi~\bar\psi(p)i\gamma_5 \tau_1\psi(p)
\right] \nonumber\\
&&~~~- \beta V \frac{\sigma^2 + \pi^2}{G} - \beta V {\cal U}[\Phi,\bar\Phi,T]~, \label{eq:LagrMFms}
\end{eqnarray}
with $A_\mu = g A_\mu^a T_a$. We introduce the mean field momentum dependent constituent quark mass $M(p)$ and
renormalized pion condensate $N(p)$:
\begin{equation}
M(p) \equiv m-2\sigma f^2(p)~,~~~N \equiv -2\pi f^2(p)~.\label{eq:mass}
\end{equation}

The thermodynamical potential $\Omega$ per unit volume in the mean field approximation can be obtained by integration
over the fermion fields in the partition function of the model, see for example Ref.~\cite{Ebert:2000pb},
\begin{eqnarray}
\Omega &=& -\left(\frac{\mu_e^4}{12\pi^2} + \frac{\mu_e^2 T^2}{6} + \frac{7\pi^2 T^4}{180}\right) + {\cal
U}[\Phi,\bar\Phi,T] + \frac{\sigma^2 + \pi^2}{G} \nonumber\\
&&~~~- T\sum_n\int \frac{d^3{\bm p}}{(2\pi)^3}~\text{Tr}~\text{log}\frac{S^{-1}(i\omega_n,{\bm p})}{T}~,
\end{eqnarray}
where the sum is over fermion Matsubara frequencies $\omega_n = \pi T(2n+1)$, and the trace is over Dirac, flavor and
color indices. The inverse quark propagator is defined as
\begin{eqnarray}
&& S^{-1}(i\omega_n,{\bm p})=\nonumber\\
&& \left(\begin{array}{cc}
  (i\omega_n+\mu-\frac{2}{3}\mu_e+iA_4)\gamma_0 -{\bm\gamma}\cdot{\bm p} -M(p) & -i\gamma_5 N(p) \\
  -i\gamma_5 N(p) & (i\omega_n+\mu+\frac{1}{3}\mu_e+iA_4)\gamma_0 -{\bm\gamma}\cdot{\bm p} -M(p)\\
\end{array}\right)\otimes{\bm 1}_c~.\nonumber\\
&&\label{eq:po}
\end{eqnarray}
Performing the trace and the sum over Matsubara frequencies we have the effective potential for $\Phi$, $\sigma$ and
$\pi$, namely
\begin{eqnarray}
\Omega &=& -\left(\frac{\mu_e^4}{12\pi^2} + \frac{\mu_e^2 T^2}{6} + \frac{7\pi^2 T^4}{180}\right) + {\cal
U}[\Phi,\bar\Phi,T] + \frac{\sigma^2 + \pi^2}{G}  -2N_c \int\!\frac{d^3\bm p}{(2\pi)^3}\left[E_+ + E_- -2p\right]\nonumber \\
&& -2 T\int\! \frac{d^3{\bm p}}{(2\pi)^3}~\text{log}\left[1+3 \Phi e^{-\beta(E_+ - \mu)} + 3\bar\Phi
e^{-2\beta(E_+ - \mu)} + e^{-3\beta(E_+ - \mu)}   \right]~\nonumber\\
&&-2 T\int\! \frac{d^3{\bm p}}{(2\pi)^3}~\text{log}\left[1+3 \Phi e^{-\beta(E_- - \mu)} + 3\bar\Phi e^{-2\beta(E_-
- \mu)} + e^{-3\beta(E_- - \mu)}   \right]~\nonumber\\
&&-2 T\int\! \frac{d^3{\bm p}}{(2\pi)^3}~\text{log}\left[1+3 \bar\Phi e^{-\beta(E_+ + \mu)} + 3 \Phi e^{-2\beta(E_+
+ \mu)} + e^{-3\beta(E_+ + \mu)}   \right]~\nonumber\\
&&-2 T\int\! \frac{d^3{\bm p}}{(2\pi)^3}~\text{log}\left[1+3 \bar\Phi e^{-\beta(E_-
+ \mu)} + 3 \Phi e^{-2\beta(E_- + \mu)} + e^{-3\beta(E_- + \mu)}   \right]~,\nonumber\\
\label{eq:O1}
\end{eqnarray}
where
\begin{equation}
E_\pm = \sqrt{(E_p \mp \mu_e/2)^2 + N^2}~,\label{eq:Epm}
\end{equation}
and $E_p = \sqrt{p^2 + M^2(p)}$. In Eq.~\eqref{eq:O1} the integral of $2p$ is an irrelevant constant that we subtract
in order to make the thermodynamical potential finite at each value of temperature and chemical potential. The ground
state of the model is defined by the values of $\sigma$, $\pi$, $\Phi$, $\bar\Phi$ that minimize $\Omega$ and that have
a vanishing total charge; the latter condition is equivalent to the requirement
\begin{equation}
\frac{\partial\Omega}{\partial\mu_e}=0~.
\end{equation}

In this paper we use the convenient Polyakov gauge,
\begin{equation}
\Phi = \frac{1}{3}\text{Tr}\left[e^{i \beta (\lambda_3 \phi_3 + i \lambda_8 \phi_8})\right]~,
\end{equation}
with $\phi_3$, $\phi_8$ real parameters. It has been widely discussed in Ref.~\cite{Roessner:2006xn} that in the mean
field approximation and with the choice of the effective potential ${\cal U}$ given by Eq.~\eqref{eq:Poly} one has
$\langle\Phi\rangle = \langle\bar\Phi\rangle$ for any value of $T$ and $\mu$, and the solution
$\langle\Phi\rangle\neq\langle\bar\Phi\rangle$ at finite $\mu$ is due to quantum fluctuations. Since in this paper we
consider only the mean field approximation we chose $\Phi = \bar\Phi$ in the calculations. This choice implies $\phi_8
= 0$ thus we are left with only one parameter $\phi_3\equiv\phi$.

Before closing this section we write the dispersion laws of the quasi-particles in the Polyakov gauge, defined as the
poles of the quark propagator given by Eq.~\eqref{eq:po}:
\begin{eqnarray}
E_{ur} = \mp\mu \pm i\phi + E_+~,&&E_{dr} = \mp\mu \pm i\phi + E_-~,\label{eq:Dr}\\
E_{ug} = \mp\mu \mp i\phi + E_+~,&&E_{dg} = \mp\mu \mp i\phi + E_-~,\label{eq:Dg}\\
E_{ub} = \mp\mu  + E_+~,&& E_{db} = \mp\mu  + E_-~.\label{eq:Db}
\end{eqnarray}
In the previous equations $u$, $d$ correspond to up and down quarks, $r$, $g$ and $b$ to the colors red, green and
blue; the upper (lower) sign multiplying $\mu$ and $\phi$ correspond to quarks (antiquarks).

\section{Susceptibilities in the PNJL model}
In order to study the landscape of the phases of the PNJL model we introduce the susceptibility matrix.
Susceptibilities are useful to identify phase transitions since they are proportional to the fluctuations of the order
parameters around their mean field values, which usually are enhanced near a phase transition. We follow closely
Ref.~\cite{Sasaki:2006ww} for the formalism settings. The first step is the definition of the dimensionless curvature
matrix of the free energy around its global minima $C$~\cite{Fukushima:2003fw,Sasaki:2006ww},
\begin{equation}
C\equiv\left(%
\begin{array}{ccc}
  C_{MM} & C_{M\Phi} & C_{M\bar\Phi} \\
  C_{M\Phi} & C_{\Phi\Phi} & C_{\Phi\bar\Phi} \\
  C_{M\bar\Phi} & C_{\Phi\bar\Phi} & C_{\bar\Phi\bar\Phi} \\
\end{array}%
\right)~.
\end{equation}
In the above equation the diagonal entries are defined as
\begin{eqnarray}
C_{MM} &=& \frac{\beta}{\Lambda}\frac{\partial^2\Omega}{\partial M^2}~, \\
C_{\Phi\Phi} = \frac{\beta}{\Lambda^3}\frac{\partial^2\Omega}{\partial \Phi^2}~,&&~~~ C_{\bar\Phi\bar\Phi} =
\frac{\beta}{\Lambda^3}\frac{\partial^2\Omega}{\partial \bar\Phi^2}~;
\end{eqnarray}
with $\beta=1/T$ and $\Lambda$ is the mass scale defining the form factor Eq.~\eqref{eq:f1}. $\Omega$ is defined in
Eq.~\eqref{eq:O1}. In what follows we denote by $M$ the constituent quark mass computed at $p=0$, which is a function
of $\mu$ and $T$.  The off diagonal entries are given by
\begin{eqnarray}
C_{M\Phi} = \frac{\beta}{\Lambda^2}\frac{\partial^2\Omega}{\partial \Phi \partial M}~,&&~~~
C_{M\bar\Phi} = \frac{\beta}{\Lambda^2}\frac{\partial^2\Omega}{\partial \bar\Phi\partial M}~,\\
C_{\Phi\bar\Phi} &=& \frac{\beta}{\Lambda^3}\frac{\partial^2\Omega}{\partial \Phi \partial\bar\Phi}~;
\end{eqnarray}
the derivatives are computed at the global minimum of $\Omega$. Notice that the proper definition of the curvature
matrix requires that we put $\Phi = \bar\Phi$, namely the mean field solution, only after differentiation.

The susceptibility matrix $\hat\chi$ is computed as the inverse of the curvature matrix $C$. We have
\begin{equation}
\hat\chi=\left(%
\begin{array}{ccc}
  \chi_{MM} & \chi_{M\Phi} & \chi_{M\bar\Phi} \label{eq:chiMM}\\
  \chi_{M\Phi} & \chi_{\Phi\Phi} & \chi_{\Phi\bar\Phi} \\
  \chi_{M\bar\Phi} & \chi_{\Phi\bar\Phi} & \chi_{\bar\Phi\bar\Phi} \\
\end{array}%
\right)~.
\end{equation}
Here $\chi_{MM}$, $\chi_{\Phi\Phi}$ and $\chi_{\bar\Phi\bar\Phi}$ denote respectively the dimensionless
susceptibilities of the constituent quark mass, of the Polyakov loop and of its complex conjugate. We also introduce
the average susceptibility
\begin{equation}
\bar\chi = \frac{1}{4}\left(\chi_{\Phi \Phi} + \chi_{\bar\Phi\bar\Phi} + 2\chi_{\Phi\bar\Phi}\right)~.\label{eq:chiAV}
\end{equation}


\section{Results and discussion}
In this Section we sketch our results. Firstly we discuss the set of parameters corresponding to the case I which
corresponds to $\bar{T}_0 = 208$ MeV. Case II is qualitatively similar to case I, therefore after the discussion of the
results obtained in the latter case we briefly show the results corresponding to the former case. Finally we compare
both qualitatively and quantitatively the cases I and III. We find that the phase structures of the models
corresponding to cases I and III are quite different.

\subsection{Case I: masses, Polyakov loop and quarkyonic matter}

\begin{figure*}[t!]
\begin{center}
\includegraphics[width=7cm]{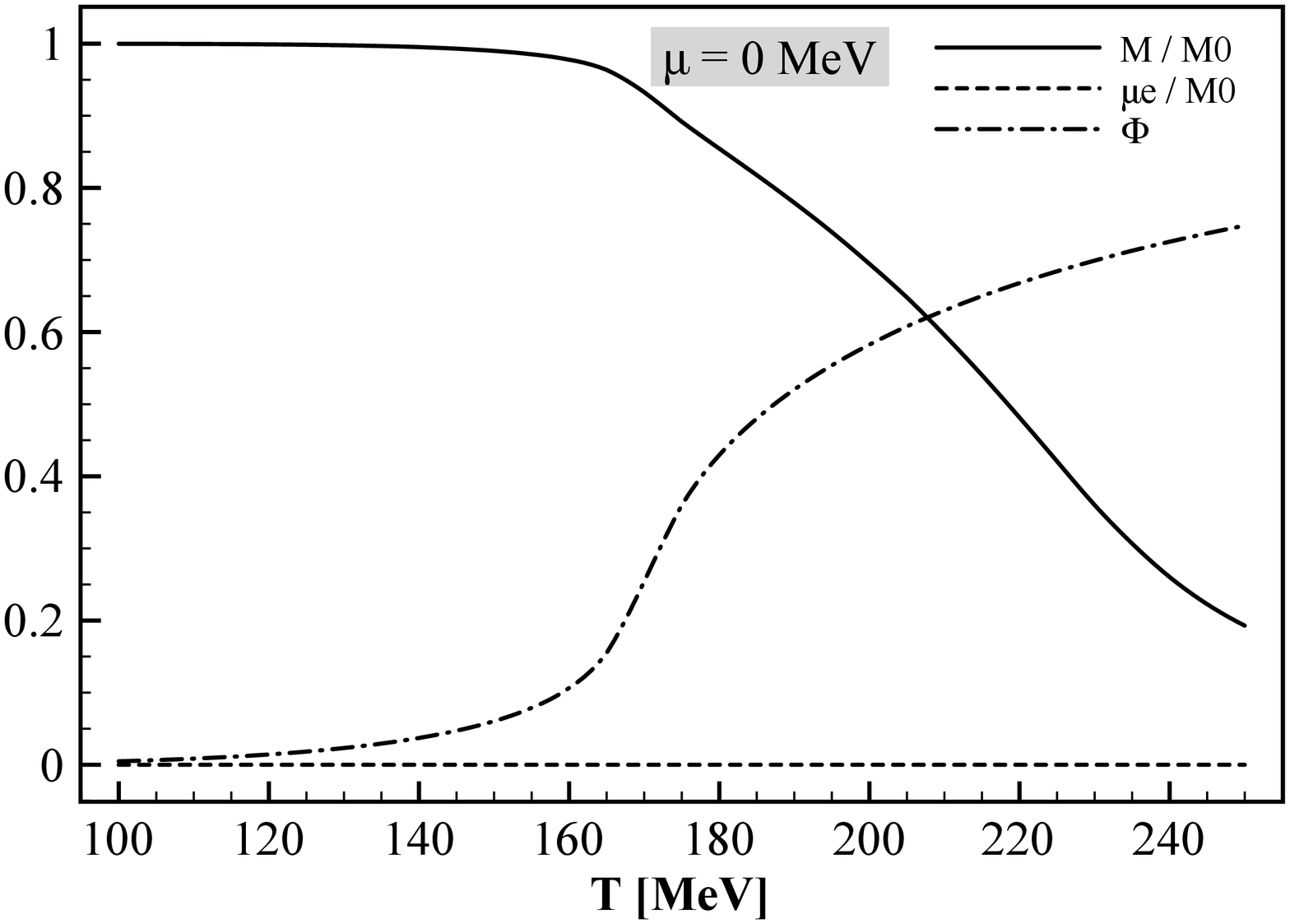}~~~~~
\includegraphics[width=7cm]{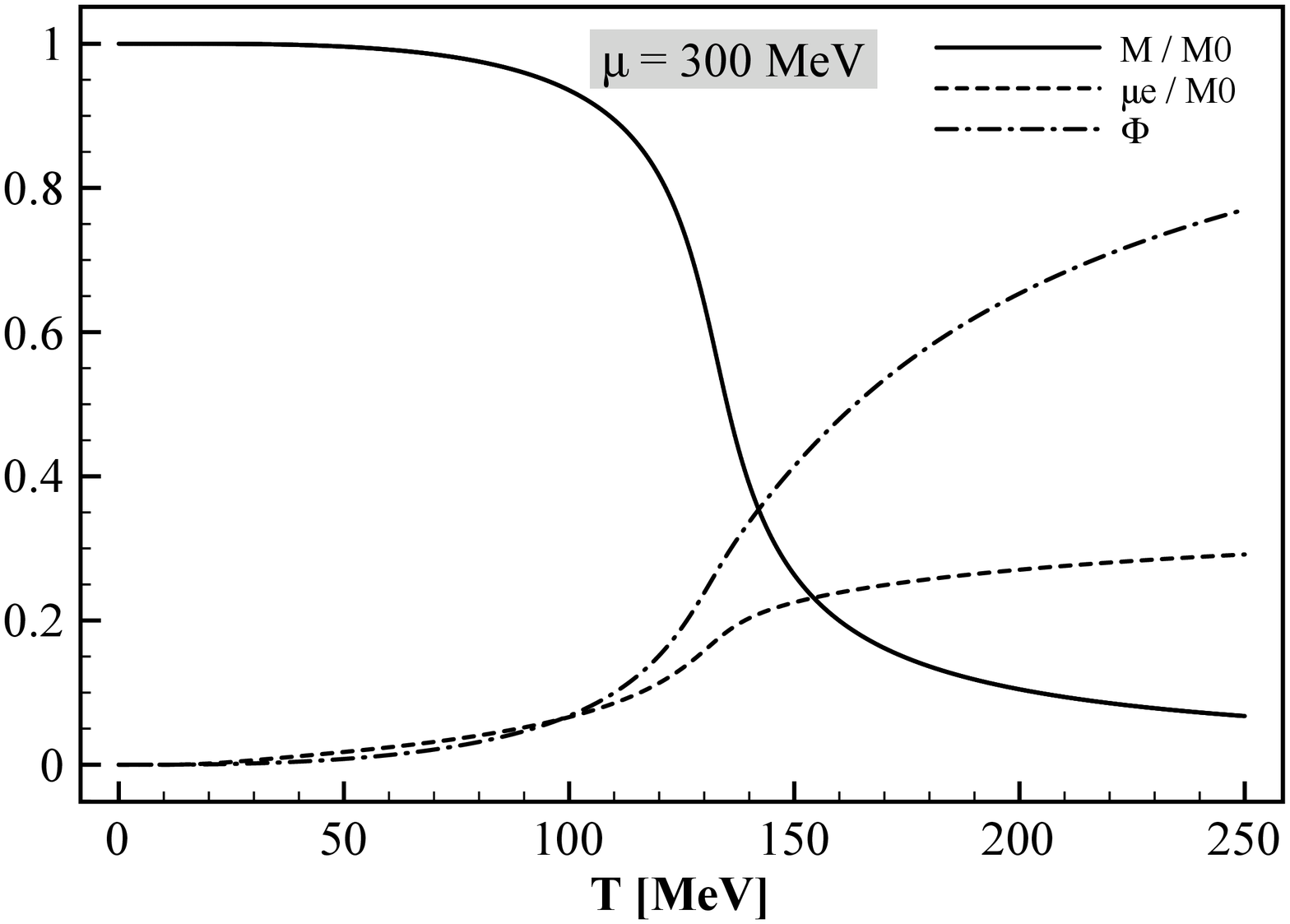}\\
\includegraphics[width=7cm]{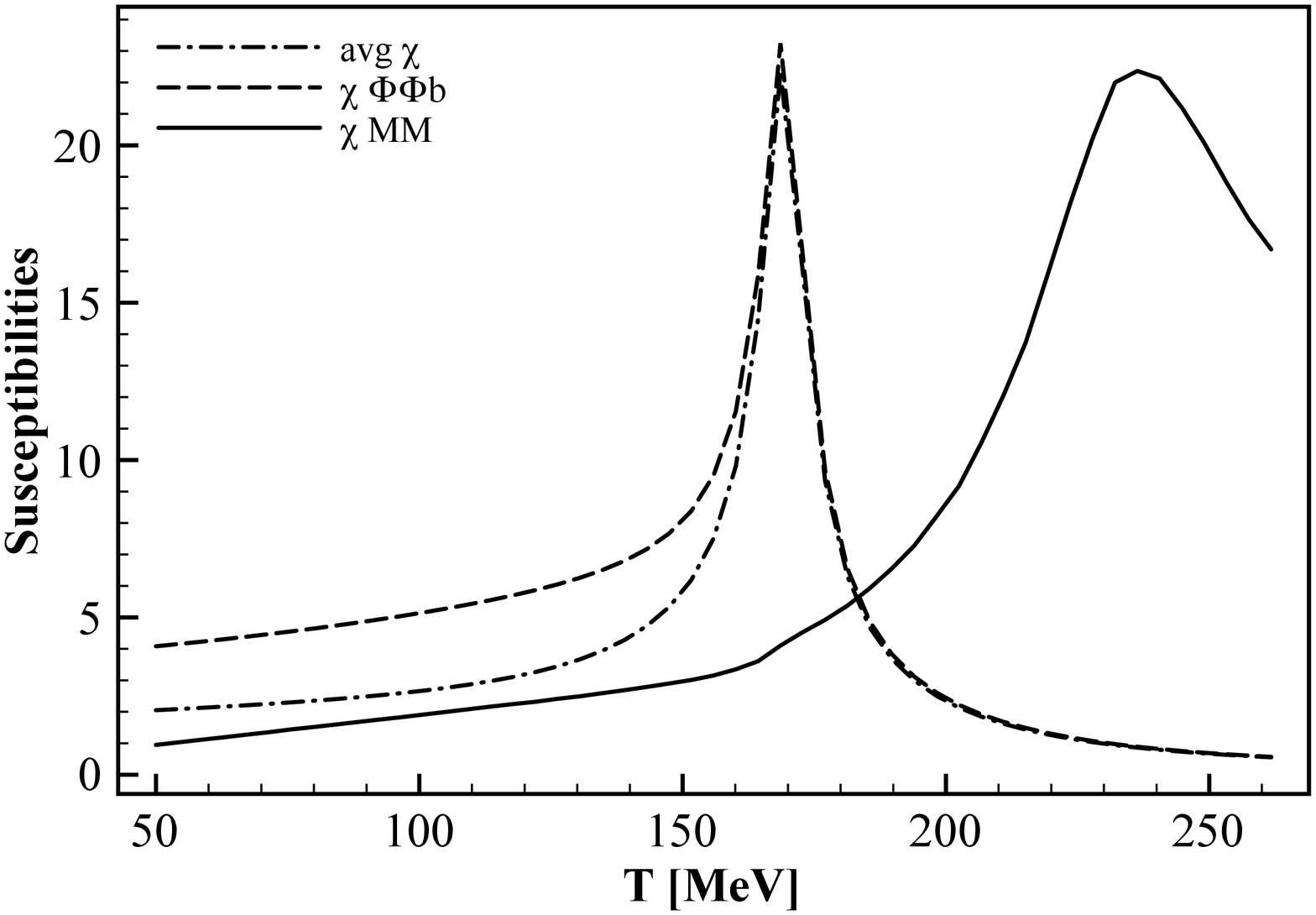}~~~~~\includegraphics[width=7cm]{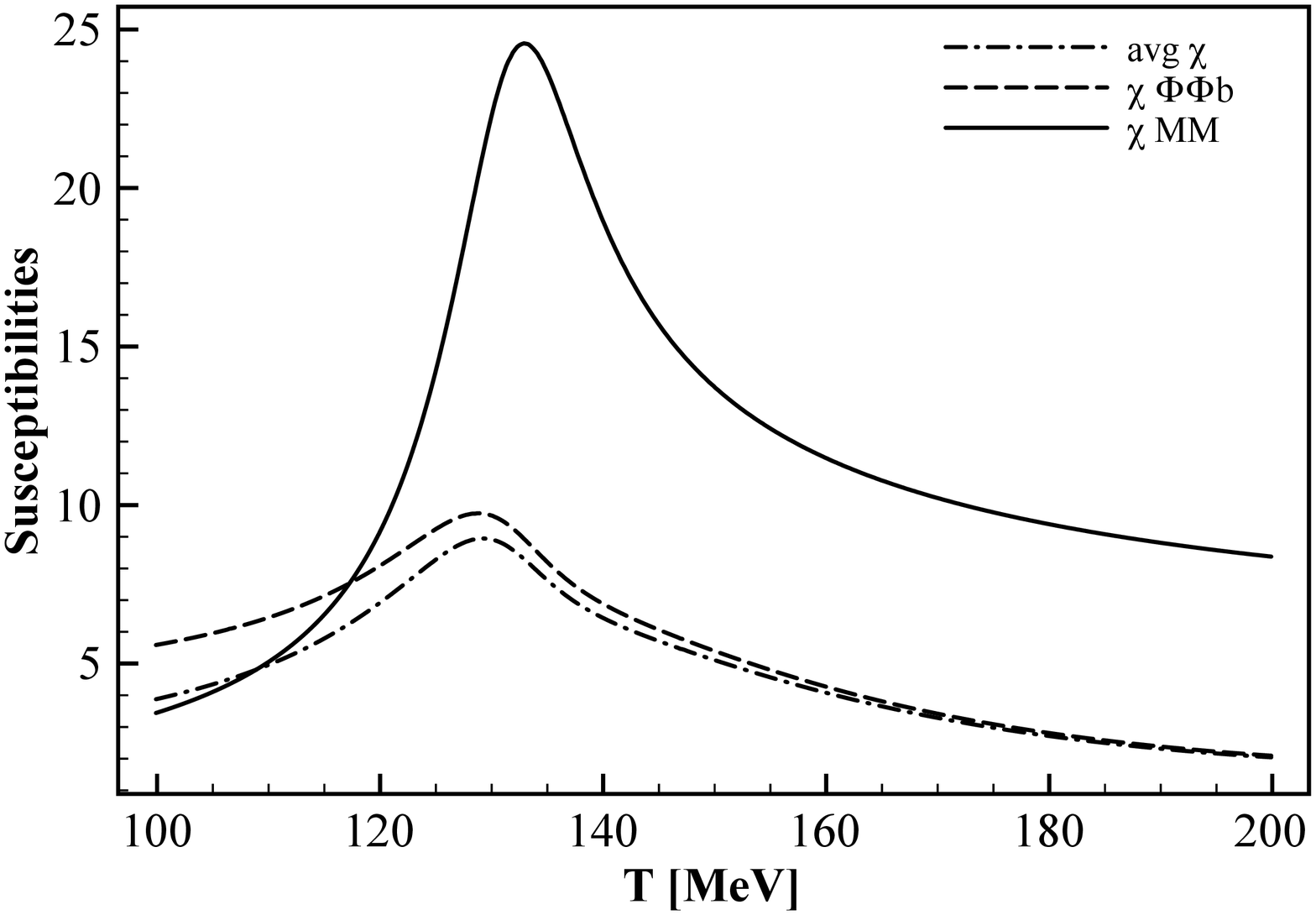}
\end{center}
\caption{\label{fig:m0} Upper panel: constituent quark mass at $p=0$, pion condensate at $p=0$ and Polyakov loop as a
function of temperature, computed at $\mu=0$ (left panel) and $\mu=353.5$ MeV (right panel). $M_0$ denotes the
constituent quark mass at $p=0$ , $\mu=0$, $\mu_e = 0$ and $T=0$, $M_0=335$ MeV. Lower panel left: susceptibilities at
$\mu=0$ as a function of temperature. Lower panel right: susceptibilities at $\mu=300$ MeV as a function of
temperature. Solid line: $\chi_{MM}$. Dashed line: $\chi_{\Phi\bar\Phi}$. Dot-dashed line: $\bar\chi$.}
\end{figure*}

In the upper panel of Fig.~\ref{fig:m0} we plot the constituent quark mass at $p=0$, the expectation value of the
traced Polyakov and the electron chemical potential as a function of the temperature, computed at $\mu=0$ (left) and
$\mu=300$ MeV (right). $M_0$ denotes the constituent quark mass at $p=0$, $\mu=0$, $\mu_e = 0$ and $T=0$, $M_0 = 335$
MeV. The pion condensate $N$ is not shown since we find $N=0$ once electrical neutrality has been imposed. The latter
result is in agreement with what we have found in our previous work, see Ref.~\cite{Abuki:2008tx}, where we have
considered the local version of the neutral two flavor PNJL model. Even if we have shown results only for two values of
the quark chemical potential, we have explicitly verified that $N$ vanishes in the whole range of chemical potentials
and temperatures considered in this work, namely $0\leq\mu\leq 500$ MeV and $0\leq T \leq 250$ MeV.

The expectation value of the Polyakov loop at $\mu=0$ is consistent with zero up to temperatures of the order of $100$
MeV.~\footnote{$\Phi$ can not be exactly zero because dynamical quarks break the $Z(3)$ symmetry explicitly;
nevertheless $\Phi$ turns out to be very small, signaling that the center symmetry is broken only softly.} It raises as
the temperature is increased becoming of the order of $1$ for temperatures close to $250$ MeV. This behavior signals a
crossover from a low temperature phase with an unbroken $Z(3)$ symmetry, to a high temperature phase with $Z(3)$
symmetry spontaneously broken. The behavior of $\Phi$ as a function of the temperature is observed even at higher
values of $\mu$, see for example the right upper panel of Fig.~\ref{fig:m0}. We call such a crossover as the $Z(3)$
crossover throughout this paper.

In the lower panel of Fig.~\ref{fig:m0} we plot three of the susceptibilities defined in the previous Section, namely
$\chi_{MM}$ (solid line), $\chi_{\Phi\bar\Phi}$ (dashed line) and $\bar\chi$ (dot-dashed line), as a function of
temperature at $\mu=0$ (left) and $\mu=300$ MeV (right). In this work we identify the chiral crossover temperature with
the temperature where $\chi_{MM}$ is maximum. In the same way and following Ref.~\cite{Sasaki:2006ww} we define the
$Z(3)$ crossover temperature as the one corresponding to the maximum of $\bar\chi$.

\begin{figure*}[t!]\begin{center}
\includegraphics[width=7cm]{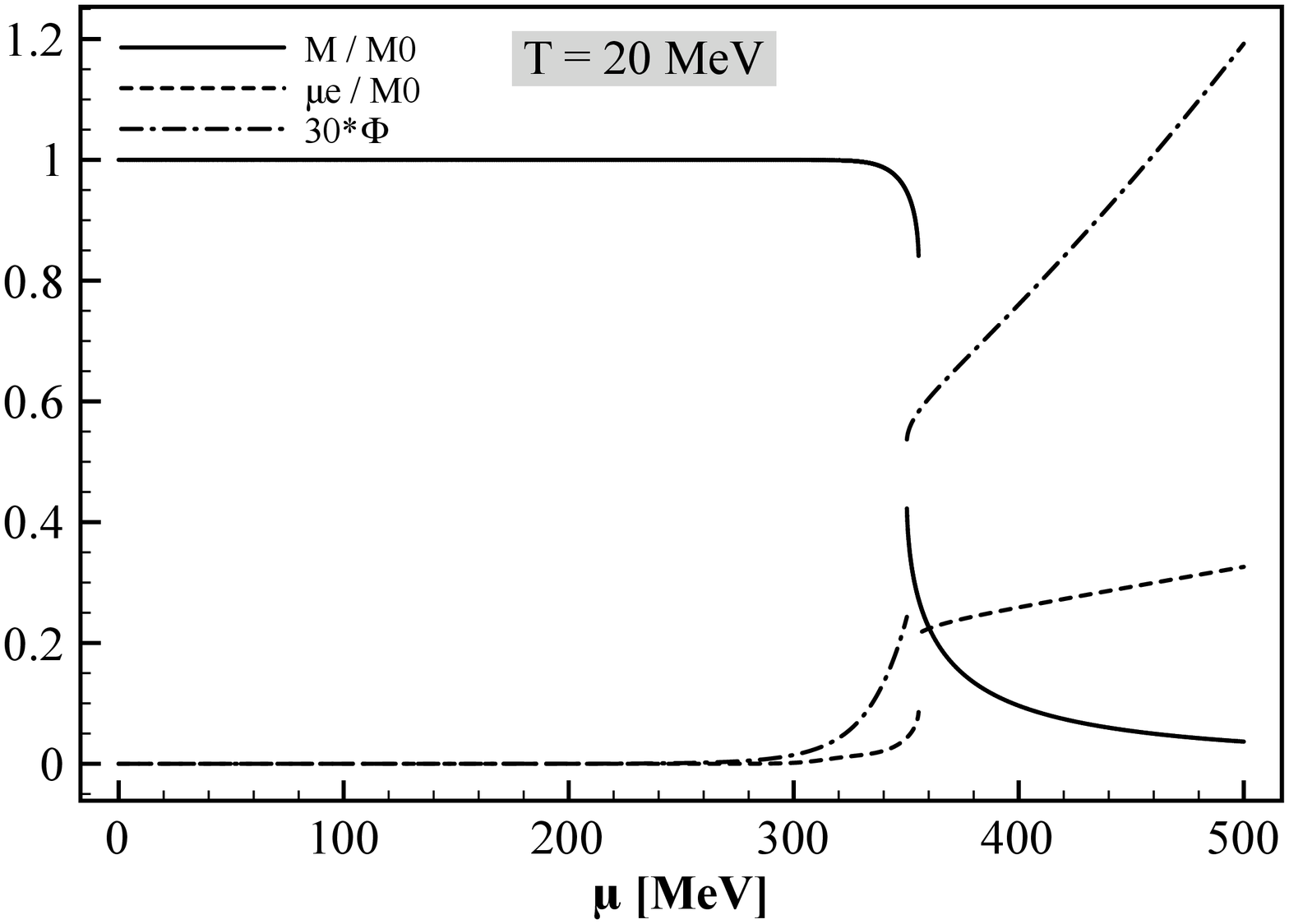}~~~~~
\includegraphics[width=7cm]{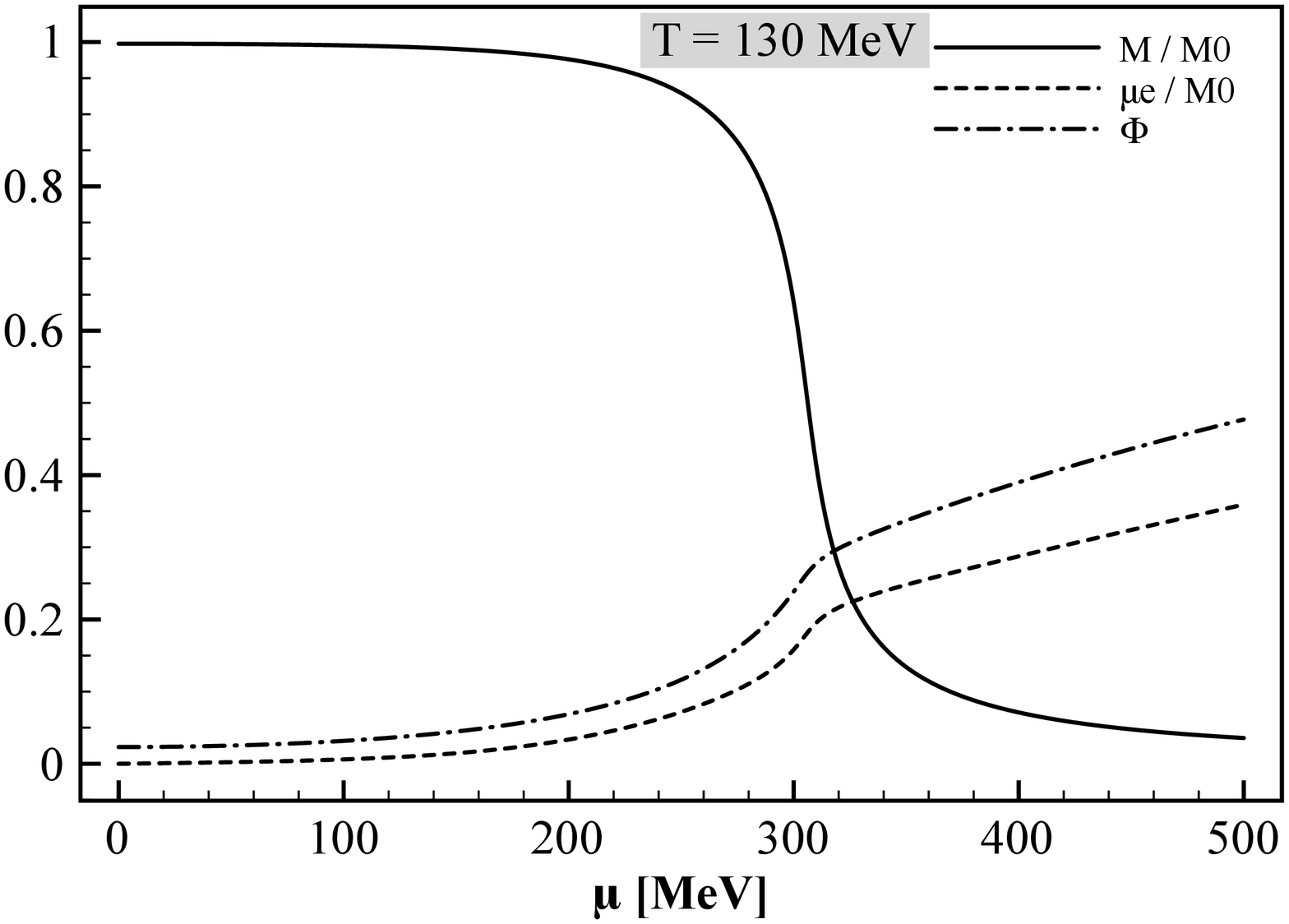}\\
\includegraphics[width=7cm]{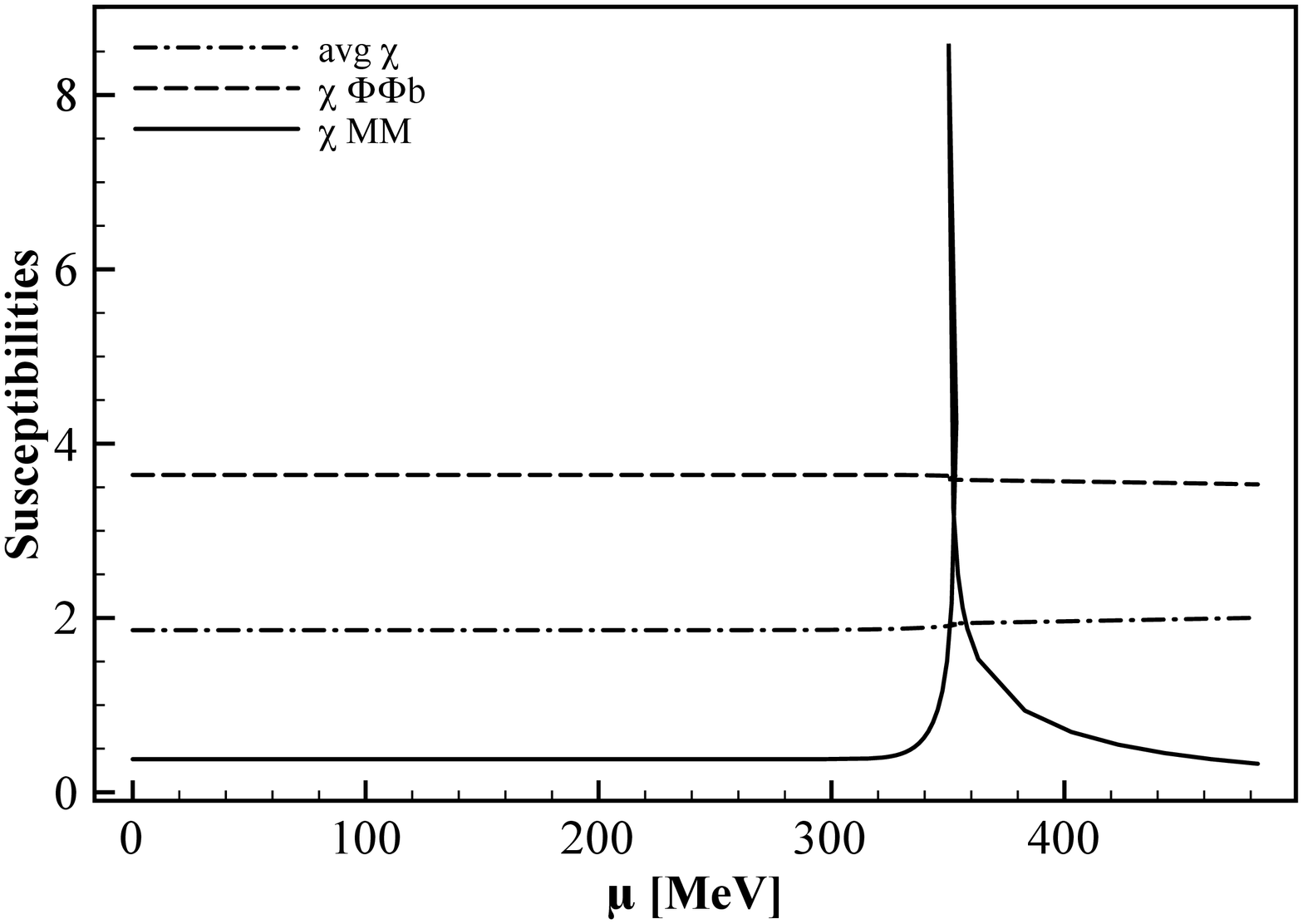}~~~~~\includegraphics[width=7cm]{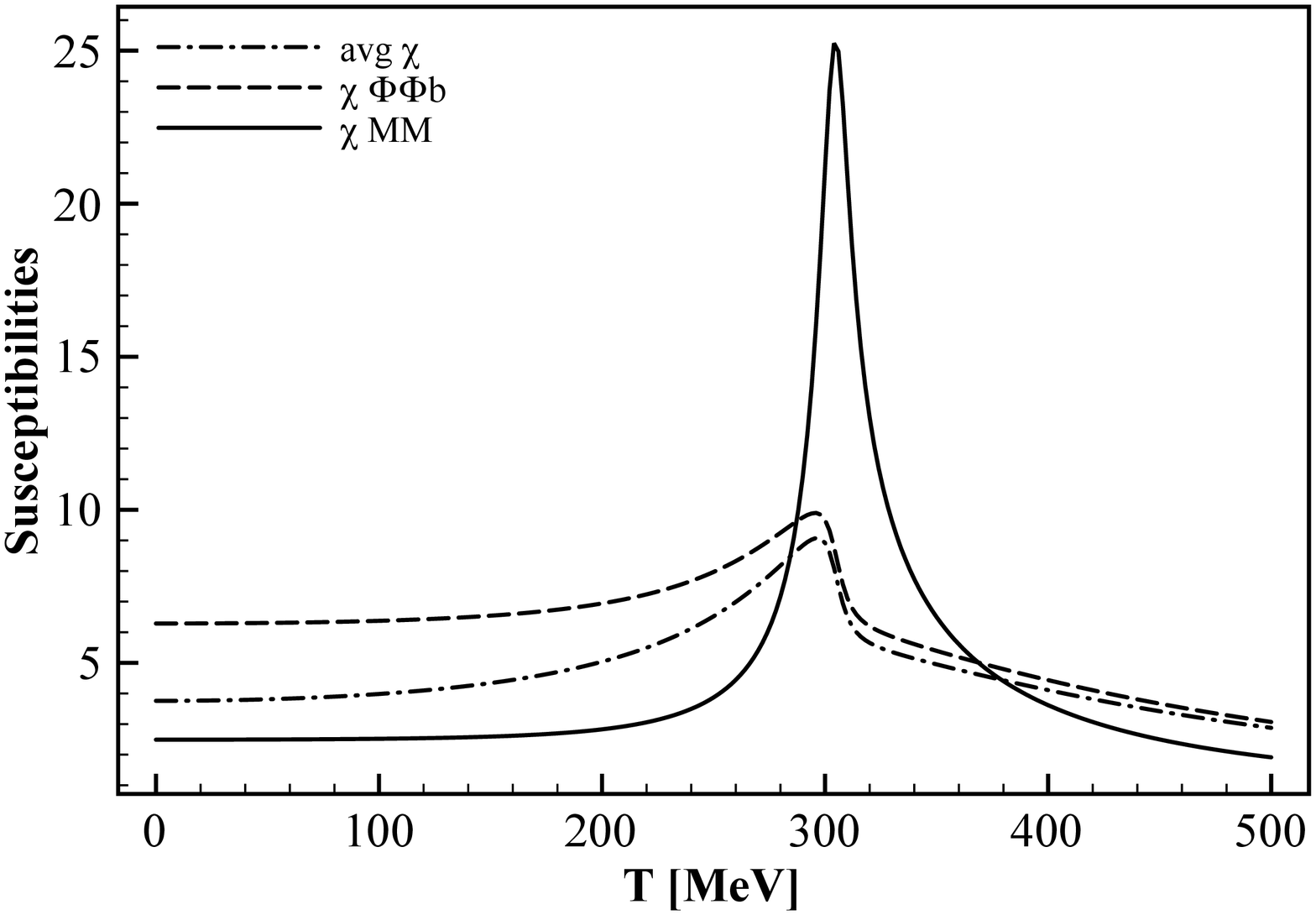}
\end{center}
\caption{\label{fig:m300} Upper panel: constituent quark mass at $p=0$, Polyakov loop and electron chemical potential
as a function of the quark chemical potential $\mu$, computed at $T=20$ MeV (left) and $T=130$ MeV (right). $M_0$
denotes the constituent quark mass at $p=0$ and $T=0$, $M_0 = 335$ MeV. In both cases $N=0$ and it is not shown. Lower
panel: susceptibilities at $T=20$ MeV (left) and $T=130$ MeV (right) as a function of the quark chemical potential.
Solid line: $\chi_{MM}$. Dashed line: $\chi_{\Phi\bar\Phi}$. Dot-dashed line: $\bar\chi$.}
\end{figure*}

We wish to investigate on the spontaneous breaking of the $Z(3)$ symmetry in the neutral PNJL model as the quark
chemical potential is increased at a fixed low temperature. To this end we plot in Fig.~\ref{fig:m300} the constituent
quark mass at $p=0$, the expectation value of the traced Polyakov and the electron chemical potential as a function of
the quark chemical potential $\mu$, computed at $T=20$ MeV (left panel).  $M_0$ denotes the constituent quark mass at
$p=0$, $\mu=0$ and $T=0$, $M_0=335$ MeV. Again we do not show the pion condensate since it turns out to vanish in the
neutral phase. At low temperatures we find a first order chiral transition at $\mu\approx 353$ MeV, in agreement with
our previous analysis~\cite{Abuki:2008tx}. In correspondence of the chiral restoration the expectation value of the
Polyakov loop has a sudden jump. Nevertheless its value remains much smaller than one even if $\mu$ is increased to
$500$ MeV, where $\Phi\approx0.04$. For comparison we show the same quantities at $T=130$ MeV in the right panel.

We now focus on the low temperature regime, therefore we refer to the left panel of Fig.~\ref{fig:m300}. In this case
we can not identify the jump of $\Phi$ as the $Z(3)$ crossover. Instead the discontinuity of $\Phi$ is simply due to
the coupling of the Polyakov loop with the chiral condensate. This is confirmed by the calculation of the Polyakov loop
susceptibilities, see the lower panel of Fig.~\ref{fig:m300}. At $T=20$ MeV, in correspondence of the jump of the
constituent quark mass, the chiral susceptibility has a pronounced peak. On the other hand the Polyakov loop
susceptibilities are very smooth functions of $\mu$ with a small cusp in correspondence of the chiral transition,
signaling the absence of a phase transition (as well as of a crossover). For comparison we show the same quantities at
$T=130$ MeV in the right panel.

\begin{figure*}[t!]\begin{center}
\includegraphics[width=7cm]{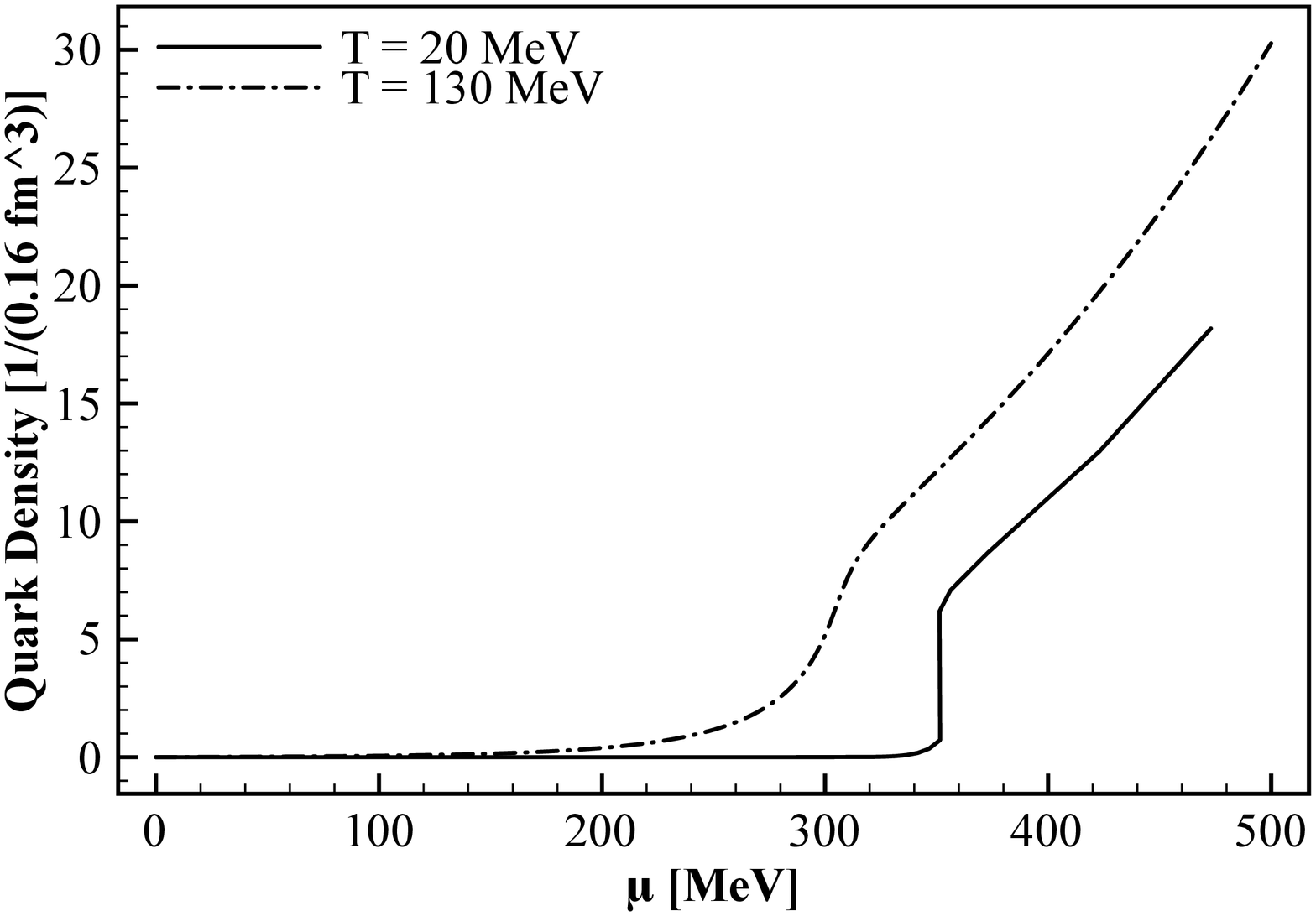}~~~\includegraphics[width=7cm]{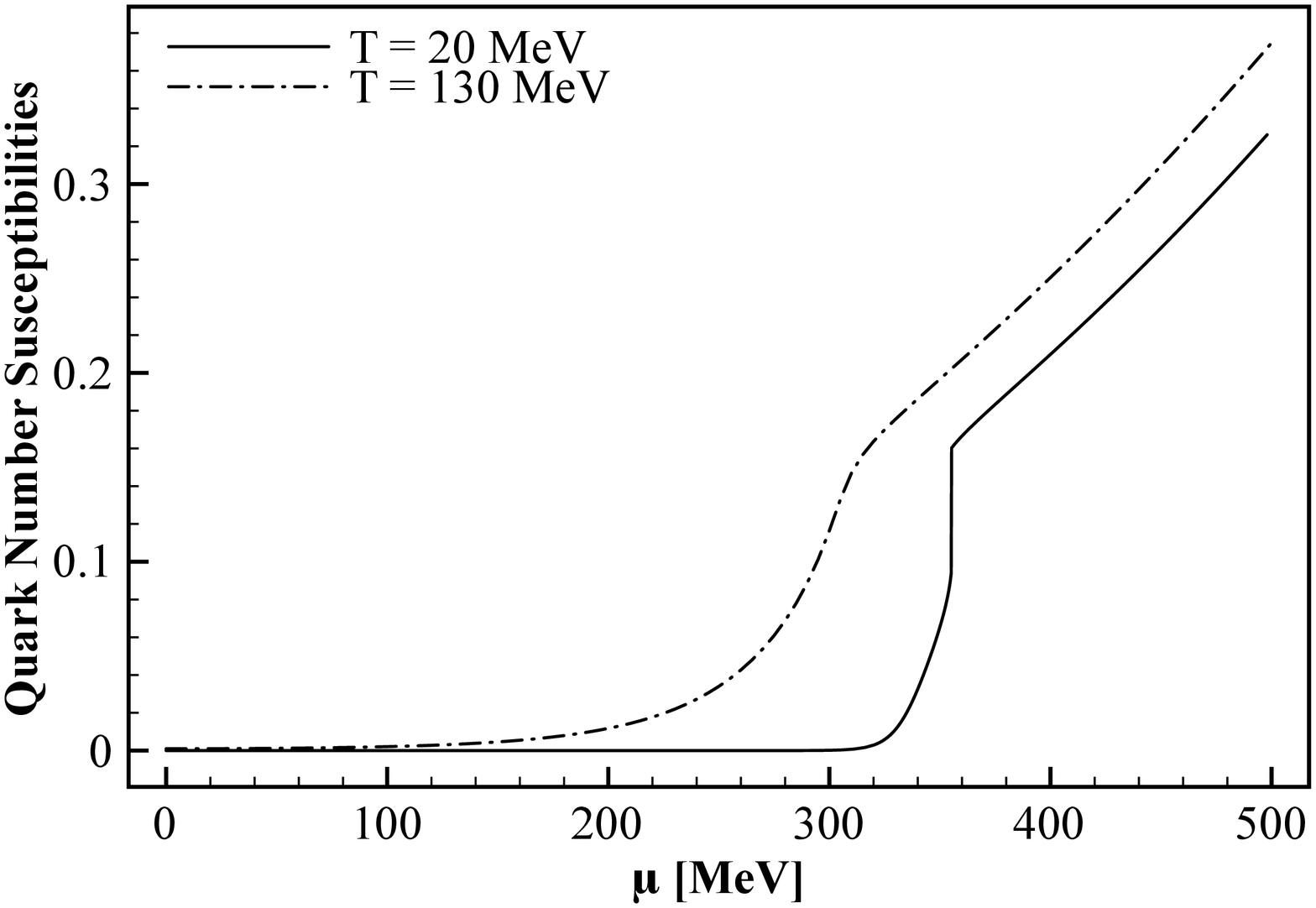}
\end{center}
\caption{\label{fig:bd} Left panel: quark number densities, in units of the nuclear saturation density $\rho_0=0.16$
fm$^{-3}$, as a function of the quark chemical potential $\mu$ at $T=20$ MeV (dashed line) and $T=130$ MeV (dot-dashed
line). Right panel: dimensionless quark number susceptibilities as a function of the quark chemical potential $\mu$ at
$T=20$ MeV (dashed line) and $T=130$ MeV (dot-dashed line).}
\end{figure*}

Our results can be interpreted by assuming that at low temperatures the $Z(3)$ symmetry is not spontaneously broken,
both at low and at high chemical potentials. The non zero value of $\Phi$ can be related to the existence of dynamical
quarks in the system, that break explicitly the center symmetry. The fact that $\Phi\ll 1$ means that in the ground
state colored quarks are suppressed (they have a finite $Z(3)$-charge), and the main contribution to the free energy is
due to the $Z(3)$-invariant multi-quark states, that are states with a zero $Z(3)$-charge. This point can be clarified
by studying the thermal population of the quasi-quarks excitations at low temperature. To this end we compute the quark
number density $n_q$,
\begin{equation}
n_q = - \frac{\partial\Omega}{\partial\mu}~,
\end{equation}
as a function of the chemical potential at fixed temperature. The result is shown in Fig.~\ref{fig:bd}. Evaluation of
the derivative of $\Omega$ defined in Eq.~\eqref{eq:O1} leads to the expression\begin{widetext}
\begin{equation}
n_q = \frac{3}{\pi^2}\int_0^\infty  p^2 dp \left[\frac{g_{+-}}{f_{+-}} + \frac{g_{--}}{f_{--}} - \frac{g_{++}}{f_{++}}
- \frac{g_{-+}}{f_{-+}}\right]~, \label{eq:n1}
\end{equation}
where we have introduced the functions
\begin{eqnarray}
f_{\pm\pm} &=& 1 + 3\Phi e^{-\beta(E_\pm \pm \mu)} + 3\Phi e^{-2\beta(E_\pm \pm \mu)} +  e^{-3\beta(E_\pm \pm \mu)}~,\\
g_{\pm\pm} &=& \Phi e^{-\beta(E_\pm \pm \mu)} + 2\Phi e^{-2\beta(E_\pm \pm \mu)} +  e^{-3\beta(E_\pm \pm \mu)}~,
\end{eqnarray}
and $E_\pm$ are defined in Eq.~\eqref{eq:Epm}. The addenda in the r.h.s. of Eq.~\eqref{eq:n1} correspond respectively
to up quarks, down quarks, up antiquarks and down antiquarks. If we put by hand $\Phi = 1$ in Eq.~\eqref{eq:n1} we
recover the usual expression of the NJL model,
\begin{equation}
n_{q,NJL} = \frac{3}{\pi^2}\int_0^\infty  p^2 dp \left[\frac{1}{1 + e^{\beta(E_+ - \mu)}} + \frac{1}{1 + e^{\beta(E_- -
\mu)}} - \frac{1}{1 + e^{\beta(E_+ + \mu)}} - \frac{1}{1 + e^{\beta(E_- + \mu)}} \right]~, \label{eq:n2}
\end{equation}
where the $3$ overall counts the number of colors. Eq.~\eqref{eq:n2} is the number density of a free fermion gas; it
shows that in the zero temperature limit and $\mu
> M$, $M$ denoting the constituent quark mass, the ground state of the NJL model is made of Fermi spheres of red, green and blue quarks.
Moreover at small but non vanishing temperatures the thermal excitations over the Fermi spheres are still quarks.

Now we compare Eq.~\eqref{eq:n2} with the analogous result of the PNJL model. At low temperature we have $\Phi \ll 1$
therefore for a rough analysis we can put $\Phi=0$ in Eq.~\eqref{eq:n1}. We are left with the expression:
\begin{equation}
n_{q,PNJL} = \frac{3}{\pi^2}\int_0^\infty  p^2 dp \left[\frac{1}{1 + e^{3\beta(E_+ - \mu)}} + \frac{1}{1 +
e^{3\beta(E_- - \mu)}} - \frac{1}{1 + e^{3\beta(E_+ + \mu)}} - \frac{1}{1 + e^{3\beta(E_- + \mu)}} \right]~.
\label{eq:n3}
\end{equation}\end{widetext}
The above equation is valid for every value of $\mu$. In the limit $T\rightarrow0$ and for $\mu > M$, with $M$ the
constituent quark mass, it gives the equation obtained in the NJL model, that is a ground state of Fermi spheres of
red, green and blue quarks at the chemical potential $\mu$. If we introduce a small temperature then the thermal
excitations are not quarks but the $Z(3)$ symmetric three quark states, that is states made of one red quark, one green
quark and one blue quark. This is clear from the above Eq.~\eqref{eq:n3} by looking at the arguments of the
exponentials in the four addenda. Each of the addenda corresponds to the occupation number of fermions with energy
given by $3E_{\pm} - 3\mu$ which is exactly the energy of the lightest $Z(3)$ symmetric state, namely (see
Eqs.~\eqref{eq:Dr}-\eqref{eq:Db})
\begin{equation}
E_{red} + E_{green} + E_{blue} = 3E_{\pm} - 3\mu~,\label{eq:E1}
\end{equation}
the sign depending on the flavor we consider ($E_+$ corresponds to up quarks, $E_-$ to down quarks). The same result
holds for antiquarks, simply by replacing $\mu\rightarrow-\mu$. The combination~\eqref{eq:E1} is exactly the argument
of the exponentials in Eq.~\eqref{eq:n3}.

To summarize: for the parametrization I the ground state of PNJL quark matter in the regime of low temperature $T\ll M$
and $\mu> M$ is made of Fermi spheres of quarks, and the thermal excitations above the aforementioned Fermi spheres are
the three quark states, neutral with respect to $Z(3)$.

For completeness, on the right panel of Fig.~\ref{fig:bd} we plot the dimensionless quark number susceptibilities,
$\chi_q$, defined as
\begin{equation}
\chi_q = -\frac{1}{\Lambda^2}\frac{\partial^2\Omega}{\partial\mu^2}~,
\end{equation}
where $\Lambda$ is the form factor momentum scale in Eq.~\eqref{eq:f1}, and $\Omega$ is the PNJL free energy given by
Eq.~\eqref{eq:O1}.

\subsection{Case I: phase diagram in the $\mu-T$ plane}
\begin{figure*}[t!]
\begin{center}
\includegraphics[width=10cm]{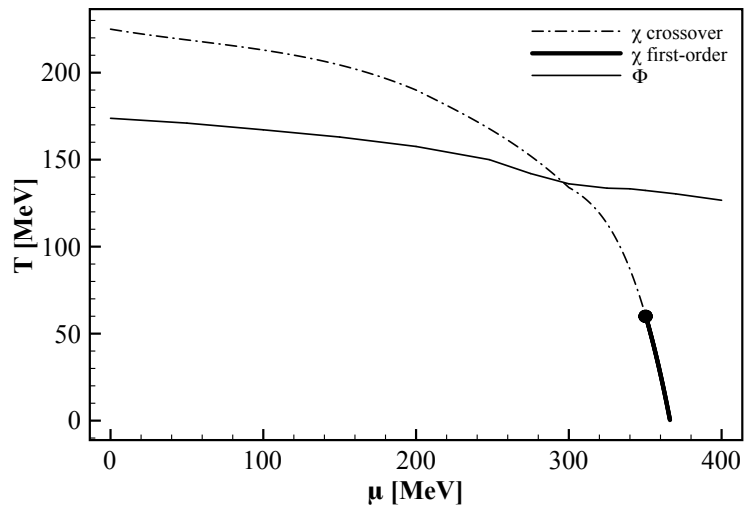}
\end{center}
\caption{\label{fig:phased} Phase diagram of the neutral two flavor PNJL model. The dot-dashed line corresponds to the
chiral crossover; the bold solid line is the first order transition. The thin solid line denotes the deconfinement
crossover.}
\end{figure*}

In Fig.~\ref{fig:phased} we summarize the phase diagram of the model in the $\mu-T$ plane with the parametrization I.
The thin line corresponds to the chiral crossover; the thick line is the first order chiral transition. We identify the
peaks (or the local maxima) in the susceptibilities with the phase transitions. In particular, the chiral crossover is
related to the peak of $\chi_{MM}$ in Eq.~\eqref{eq:chiMM}; on the other hand following Ref.~\cite{Sasaki:2006ww} we
identify the peak of the average susceptibility $\bar\chi$ defined in Eq.~\eqref{eq:chiAV} with the Polyakov loop
crossover.

From the qualitative point of view the phase diagram does not differ from our previous result~\cite{Abuki:2008tx}
obtained in the sharp cutoff regularization scheme. The chiral crossover at $\mu = 0$ is located at $T_c = 215$ MeV, to
be compared with our previous work~\cite{Abuki:2008tx} $T_c = 206$ MeV. The critical end point is only slightly
shifted: in this work we find
\begin{equation}
(\mu_E,T_E)\approx(350,55)~\text{MeV}~;
\end{equation}
this result has to be compared with~\cite{Abuki:2008tx}
\begin{equation}
(\mu_E,T_E)\approx(340,80)~\text{MeV}~,~~~\text{sharp cutoff.}
\end{equation}
Finally at $T=0$ we find that the chiral crossover occurs at $\mu = 370$ MeV, while in our previous work we have found
$\mu = 350$ MeV.

We now discuss the Polyakov loop crossover line, corresponding to the thin solid line in Fig.~\ref{fig:phased}. At
small values of the quark chemical potential the peaks of the averaged susceptibility are well pronounced, see for
example Fig.~\ref{fig:m0}. As $\mu$ is increased the peaks of $\bar\chi$ as well as of the diagonal $\chi_{\Phi\Phi}$,
$\chi_{\bar\Phi \bar\Phi} $ and off-diagonal $\chi_{\bar\Phi \Phi}$ susceptibilities are broadened and the crossover is
dilute over a wide interval of temperatures, see the right panel in Fig.~\ref{fig:m0}. In the window of chemical
potential studied in this paper, $0\leq \mu\leq 500$ MeV, we are still able to observe maxima of $\bar\chi$ (as well as
for the other susceptibilities) as a function of the temperature at a fixed value of $\mu$; the width of the maxima
increases as $\mu$ is increased. Therefore we expect that at high values of $\mu$ and $T$ the peaks of $\bar\chi$ will
be very dilute, meaning that the crossover disappears in the model under consideration. This result changes if we
consider $\mu$ dependent coefficients of the Polyakov loop effective potential as we discuss later.

We finally notice that our results for the $Z(3)$ crossover is in qualitative agreement with the results obtained in
Ref.~\cite{Sasaki:2006ww}, where the authors study the phase diagram and the susceptibilities of the PNJL model with
quarks at the same chemical potential, and with a polynomial form of the Polyakov loop effective potential ${\cal U}$.
This suggests that the $Z(3)$ crossover is not mainly governed by the specific form of ${\cal U}$ or by electrical
neutrality, but by the assumption that the deconfinement scale $\bar{T}_0$ in ${\cal U}$ is kept independent on $\mu$
in this calculation.

\subsection{Case II: critical points}
From the qualitative point of view the case with $\bar{T}_0 = 270$ MeV does not differ from the previously analyzed
case II. Therefore we simply give the coordinates of the critical points obtained in this case. At $\mu=0$ we find the
chiral crossover at $T=219$ MeV and the $Z(3)$ crossover at $T=211$ MeV. The critical end point coordinates are
\begin{equation}
(\mu_E,T_E)\approx(336,103)~\text{MeV}~,~~~\bar{T}_0=270~\text{MeV}~.
\end{equation}


\subsection{Case III: critical points and phase structure}
\begin{figure*}[t!]\begin{center}
\includegraphics[width=7cm]{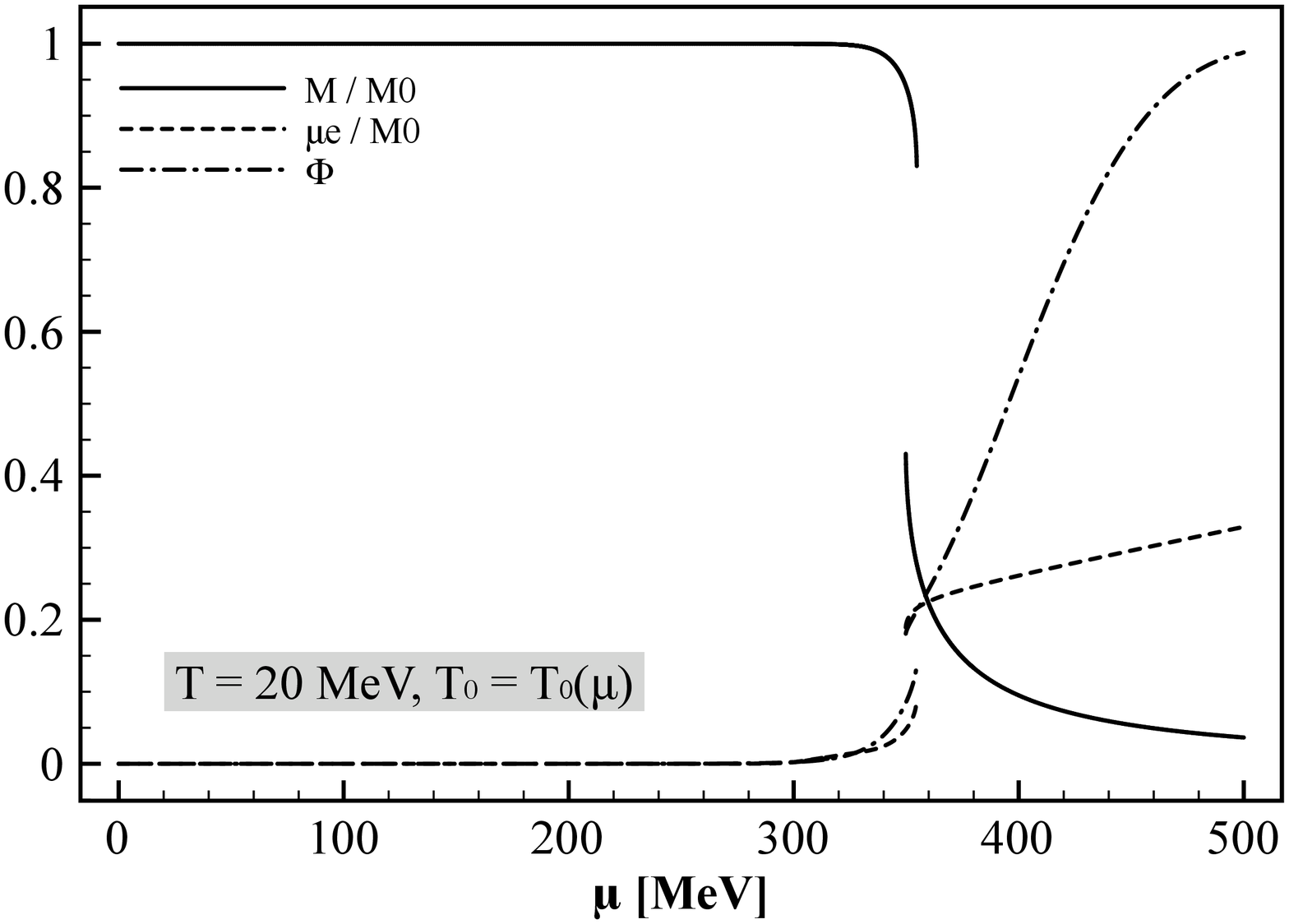}~~~~~
\includegraphics[width=7cm]{hiro1.eps}\\
\includegraphics[width=7cm]{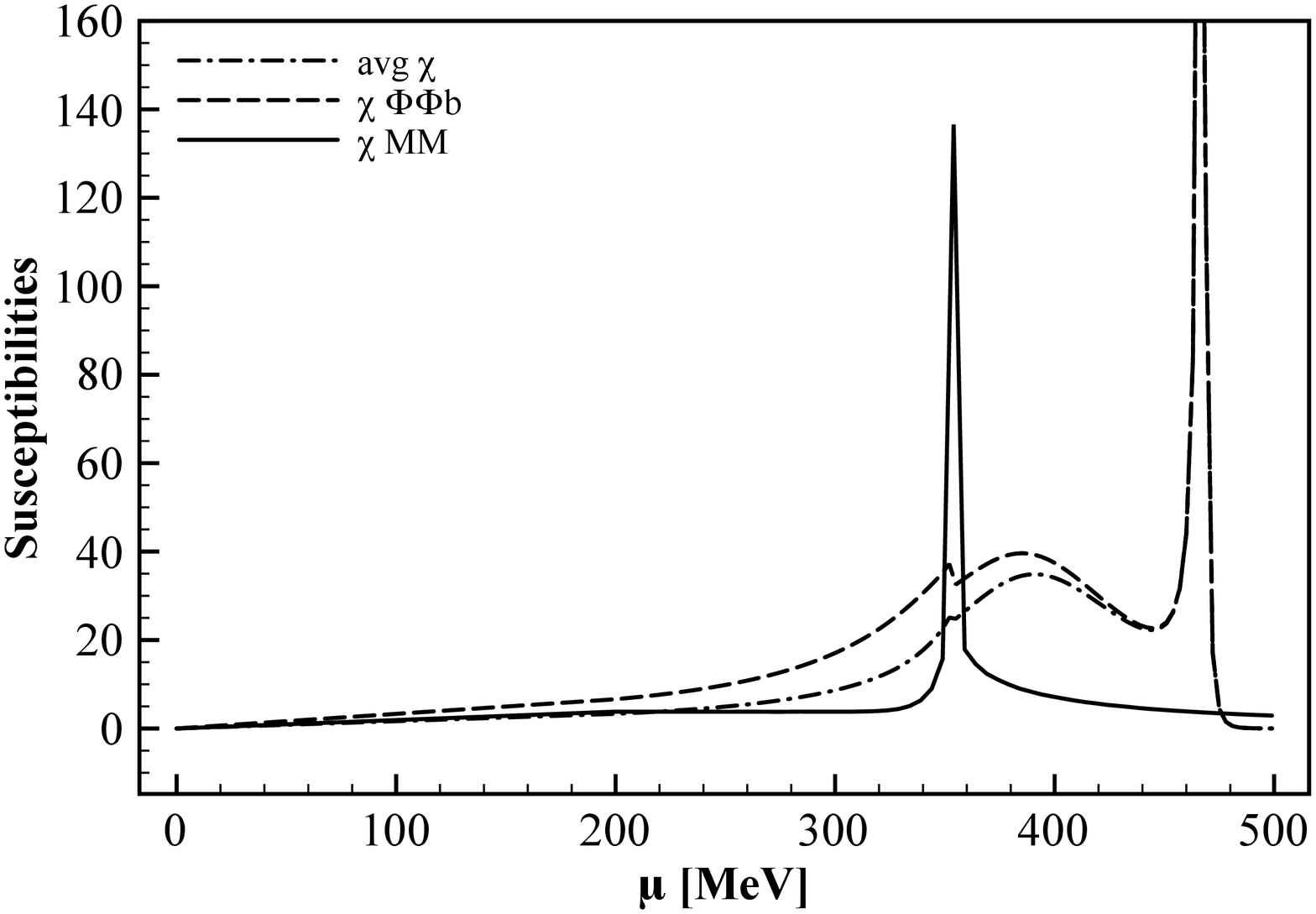}~~~~~\includegraphics[width=7cm]{ergoMU20.eps}
\end{center}
\caption{\label{fig:caxx1} Upper panel: constituent quark mass at $p=0$, Polyakov loop and electron chemical potential
as a function of the quark chemical potential $\mu$, computed at $T=20$ MeV in the case III (left) and case I (right,
shown for comparison with case III; it is the same plot shown in Fig.~\ref{fig:m300}). $M_0$ denotes the constituent
quark mass at $p=0$, $\mu=0$ and $T=0$, $M_0 = 335$ MeV. In both cases $N=0$ and it is not shown. Lower panel:
susceptibilities at $T=20$ MeV in the case III (left) and case II (right). Solid line: $\chi_{MM}$. Dashed line:
$\chi_{\Phi\bar\Phi}$. Dot-dashed line: $\bar\chi$.}
\end{figure*}

We now discuss the results obtained in the case III in which we assume both a $\mu$ and a $N_f$ dependence of the
parameter $\bar{T}_0$ of the Polyakov loop potential, see Eq.~\eqref{eq:T0m}. Our main goal is to emphasize the
differences between case III and case I. The main difference arises at low temperature and high chemical potential, so
we focus on this regime. In Fig.~\ref{fig:caxx1} we plot in the left panel the constituent quark mass at $p=0$ and the
expectation value of $\Phi$ as a function of $\mu$ at $T=20$ MeV, with the related susceptibilities, for the case III,
and compare these results with those obtained in the case I at the same temperature (right panel). We have verified
that qualitatively the picture does not change if we lower the temperature to the order of one MeV.

At $\mu=0$ the critical temperatures are equal to those computed in case I (simply because $\bar{T}_0(\mu=0) = 208$
MeV). Moreover the coordinates of the critical end point are
\begin{equation}
(\mu_E,~T_E)=(339,53)~\text{MeV}~,~~~\bar{T}_0=\bar{T}_0(\mu)~.
\end{equation}

The data on $\Phi$ corresponding to the parametrization III show that the case $\bar{T}_0 = \bar{T}_0(\mu)$ is quite
different from the case $\bar{T}_0=208$ MeV. In the case III (left panel) in correspondence of the chiral transition at
$\mu\equiv\mu_c\approx 350$ MeV the Polyakov loop has a net jump from $\Phi\ll 1$ at $\mu=\mu_c - 0^+$ to a definitely
non zero value $\Phi\approx 0.3$ at $\mu=\mu_c + 0^+$. Since the contribution of the one and two quark states ($Z(3)$
charges) to the free energy is multiplied by $3\Phi$, see Eq.~\eqref{eq:O1}, and in the present case $3\Phi$ is of the
order of unity, the weight of the $Z(3)$ charges in the free energy is the same of the weight of the three quark
states. This behavior is different from what we have found in the case of $\bar{T}_0 = 208$ MeV. The similarity between
the two cases is partially recovered if we consider temperatures of the order of one MeV; in this case we find a narrow
window in $\mu$ where $3\Phi$ is of the order of $0.1$, revealing a ground state in which the leading contribution to
the free energy comes from the thermal excitations of $Z(3)$ neutral states. We discuss this point in more detail in
the following Section. Finally the analysis of the peaks of the susceptibilities $\chi_{\bar\Phi\Phi}$ and $\bar\chi$
(lower left panel in Fig.~\ref{fig:caxx1}) reveals that the $Z(3)$ crossover occurs at $\mu\approx 460$ MeV.

\section{Comparison between the two scenarios}
In this Section we compare the qualitative picture that arises from the study of the phase diagram of the neutral PNJL
model within two scenarios: the first one corresponds to keeping an independent $\bar{T}_0$, case I; the second one
corresponds to keeping a $\mu$-dependent $\bar{T}_0$,  case III.

The results that we have discussed in the previous Sections show that the phase diagram of the PNJL model in the case I
at low temperatures is similar to the phase diagram obtained in the large $N_c$ approximation of QCD, see
Refs.~\cite{McLerran:2007qj,Hidaka:2008yy}. At low temperatures the latter phase diagram consists of two regions: the
first one at low values of $\mu$, defined as the {\it confined phase} and characterized by $\Phi=0$ and a vanishing
baryon density; the second one at large values of $\mu$ called {\it quarkonia} in which $\Phi=0$ but the baryon density
is not vanishing. Finally at high temperature one finds the {\it deconfined phase} with $\Phi\neq 0$ and a non
vanishing baryon density. In the quarkyonic phase the free energy is that of free quarks, but the thermal excitations
are those of baryons. Our previous discussion and Figs.~\ref{fig:m300} and~\ref{fig:bd} show that this happens even in
the PNJL model in the low temperature regime. Therefore the PNJL model with parametrization I approximately reproduces
the large $N_c$ phase diagram at low temperatures, if one interprets the state with $\Phi\ll 1$ at high $\mu$ with the
quarkyonic phase of large $N_c$. This fact has been already noticed in a study of the three flavor model by
Fukushima~\cite{Fukushima:2008wg} where the author has suggested to identify the low temperature-high density ground
state of the model as the quarkyonic phase of large $N_c$ QCD. Our results strengthen this idea and thus suggest that
the quarkyonic-like ground state of low temperature-high density PNJL model is not a peculiarity of the three flavor
case, but it seems to be a characteristic of the PNJL model itself, as far as we do not include an explicit $\mu$
dependence into the coefficients of ${\cal U}$ (we discuss this case in a next Subsection). The main difference between
large $N_c$ and PNJL is that in the latter model one can excite one and two quark states (that is $Z(3)$ charges) if
the temperature is high enough. As a consequence, the deconfinement transition observed in the large $N_c$ model at
high temperature and high chemical potential is replaced in the present model by a smooth $Z(3)$ crossover.

\begin{figure*}[t!]
\begin{center}
\includegraphics[width=10cm]{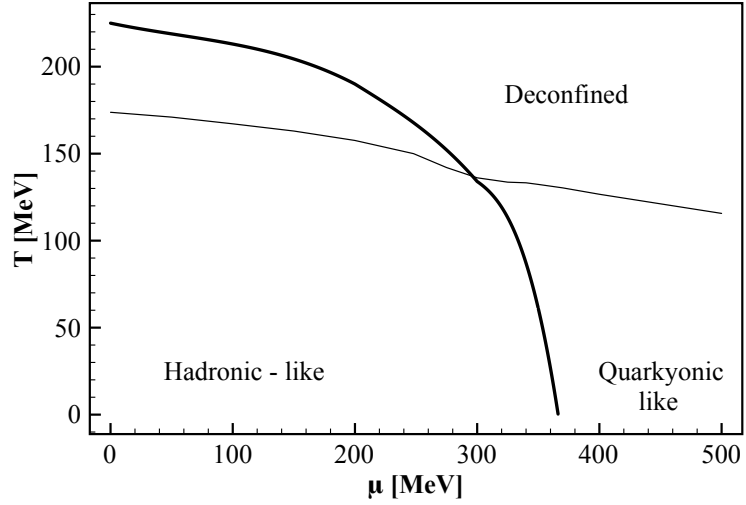}
\end{center}
\caption{\label{fig:phase-im} Cartoon phase diagram of the neutral two flavor PNJL model and comparison with that
obtained in the large $N_c$ approximation~\cite{McLerran:2007qj}. The bold line denotes the chiral crossover as well as
the chiral first order transition. The thin line corresponds to the deconfinement crossover. Both of these lines are
the same which we have shown in Fig.~\ref{fig:phased}. The $n_q$ crossover coincides with the chiral one. At low
temperature we have $\Phi\approx0$ and $n_q\approx 0$ in the chiral broken phase,  in agreement with the hadronic phase
of Ref.~\cite{McLerran:2007qj}. At low temperature and in the chiral symmetric phase we find $\Phi\approx0$ and
$n_q\neq0$ in agreement with the quarkyonic phase~\cite{McLerran:2007qj}. For these reasons we have called the two low
temperature regions {\em hadronic-like} and {\em quarkyonic-like} respectively.}
\end{figure*}

In Fig.~\ref{fig:phase-im} we show a cartoon phase diagram of the neutral two flavor PNJL model and a comparison with
that obtained in the large $N_c$ approximation~\cite{McLerran:2007qj}. The bold line denotes the chiral crossover as
well as the chiral first order transition. The thin line corresponds to the deconfinement crossover. Both of these
lines are the same which we have shown in Fig.~\ref{fig:phased}. Since this is simply a cartoon we do not distinguish
between the crossover (small $\mu$) and first order transition (higher values of $\mu$). In the PNJL model the quark
density does not vanish at any finite temperature, even if for small chemical potential $n_q$ is very small at low
temperature (see Fig.~\ref{fig:bd}). To compare the phase diagram of the PNJL model with that of the large $N_c$
approximation we need a criterion to say if $n_q$ is zero or not. Analogously to Ref.~\cite{Fukushima:2008wg} we
identify the $n_q$ crossover with the value of $\mu$ corresponding to the inflection point of the quark density. We
find that the $n_q$ crossover defined in this way coincides with the chiral crossover as in
Ref.~\cite{Fukushima:2008wg}. Therefore the chiral crossover line in Fig.~\ref{fig:phase-im} represents the density
crossover as well. In the chiral broken phase and at low temperature $n_q\approx0$. On the other hand $n_q\neq0$ in
correspondence to the chiral symmetric phase at low temperature. At high temperature $n_q\neq0$ both in the chiral
broken and in the chiral restored phases.

At low temperature we have $\Phi\approx0$ both on the left and on the right of the dashed line, see Figs.~\ref{fig:m0}
and~\ref{fig:m300}. At low temperature the region with broken chiral symmetry has the same characteristics of the
hadron phase found in Ref.~\cite{McLerran:2007qj}; on the other hand at low temperature the region on the right of the
dashed line has the same characteristics of the quarkyonic phase found in Ref.~\cite{McLerran:2007qj}. For these
reasons we have called the two regions {\em hadronic-like} and {\em quarkyonic-like} respectively. We stress that this
analogy holds strictly speaking only at low temperature (for temperatures of the order of one hundred MeV $n_q\neq0$
even in the chiral broken phase, see Fig.~\ref{fig:bd}). Finally at high temperature (above the $Z(3)$ transition line)
we have both $\Phi$ of order of unity and $n_q\neq 0$. In analogy to the terminology of Ref.~\cite{McLerran:2007qj} we
call this region of the phase diagram the {\em deconfined-like} phase.

We briefly compare the results discussed above in relation with the case I with those obtained in the large $N_c$
approximation and at $T=0$ in Ref.~\cite{Glozman:2008kn}, where the author discusses a gap in the spectrum of
quarkyonic matter within a model. Such a gap is given by the pion mass, $M_\pi$, which becomes larger as $\mu$ is
increased. Even if the values of $M_\pi$ as a function of $\mu$ computed in Ref.~\cite{Glozman:2008kn} might differ
from the non local PNJL ones, the calculations of $M_\pi$ carried out in Refs.~\cite{Hansen:2006ee,Abuki:2008tx} using
the local NJL model show that the qualitative behavior of $M_\pi$ as a function of $\mu$ is the same in the two models.
Thus in the PNJL model we expect a large pion mass at large $\mu$ as well. However this mass does not correspond to the
gap in the excitations spectrum in our model. As a matter of fact in the quarkyonic-like region of the phase diagram in
Fig.~\ref{fig:phase-im} the three quark states can be excited; each quark has a constituent mass $M(p)$ given by
Eq.~\eqref{eq:mass} and plotted in Figs.~\ref{fig:m0} and~\ref{fig:m300}, hence the three quark state has a mass
$3M(p)$ which at small quark momenta and large $\mu$ is of the order of $10$ MeV. Therefore in our case a gap in the
spectrum still exists but it is given by the three quark state mass which is much lighter than $M_\pi$.

\begin{figure*}[t!]
\begin{center}
\includegraphics[width=10cm]{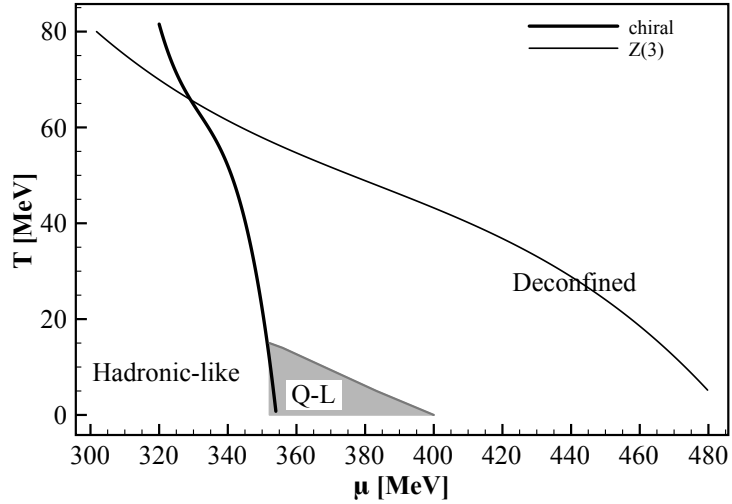}
\end{center}
\caption{\label{fig:phasedCOMP} Low temperature phase diagram of the PNJL model with parametrization III. Bold line
denotes the chiral transition. Thin line corresponds to the $Z(3)$ transition. The shaded region denoted by Q-L
corresponds to the zone of the quarkyonic-like state of matter.}
\end{figure*}

We now turn on the parametrization III. As discussed in the previous Section, the small chemical potential region of
the phase diagram in case III is qualitatively similar to that obtained in the case I, therefore we focus on the low
temperature/large chemical potential region from now on. In Fig.~\ref{fig:phasedCOMP} we draw the low temperature phase
diagram of the PNJL model with parametrization III. The bold line denotes the chiral transition; the thin line
corresponds to the $Z(3)$ transition. As in the previous section the transition lines are computed by looking at the
peaks of the chiral and $\bar\chi$ susceptibilities. The diagram in Fig.~\ref{fig:phasedCOMP} should be compared with
the analogous diagram obtained for the parametrization I which is shown in Fig.~\ref{fig:phased}. The main effect of
choosing the parameter $\bar{T}_0$ as a $\mu$ dependent one in the Polyakov loop potential is the lowering of the
$Z(3)$ transition line. Moreover the wide quarkyonic-like window in Fig.~\ref{fig:phase-im} is shrunk to a small region
in Fig.~\ref{fig:phasedCOMP}. At low temperature it is enough to reach a chemical potential of the order of $500$ MeV
to have $\Phi\approx 1$ and a net quark density; both these characteristics define the deconfined phase of
Fig.~\ref{fig:phase-im}~\cite{McLerran:2007qj}. Even if we have used the particular form $\bar{T}_0(\mu)$ suggested in
Ref.~\cite{Schaefer:2007pw} we are confident that the aforementioned results are simply due to the lowering of the
deconfinement scale $\bar{T}_0$ as $\mu$ is increased and not to the detailed analytical form of $\bar{T}_0(\mu)$. Thus
our picture should be qualitatively robust.

Before closing this Section we make a brief comment on the possible study of the scenarios discussed above on lattice.
Recently the density of states (DOS) method has been used to investigate the QCD phase transition at large
$\mu$~\cite{Fodor:2007vv}. In this paper the QCD phase diagram is mapped by studying the plaquette expectation value in
the $\mu-T$ plane. Although the lattice size implemented in~\cite{Fodor:2007vv} is relatively small and a finite volume
study is still missing, so that the results should be taken as preliminary, an interesting phase transition is observed
as $\mu$ crosses a critical value $\mu_c$ at a fixed temperature. Moreover the quark number shows a sudden rise as
$\mu$ reaches $\mu_c$. The qualitative behavior is similar in the PNJL model, see Figs.~\ref{fig:m0},~\ref{fig:m300}
and~\ref{fig:bd}. In the PNJL calculation with parametrization of Case I and II (fixed values of $T_0$) at low
temperature and in the correspondence of the the chiral crossover a small jump of the Polyakov loop occurs, the true
$Z(3)$ crossover being shifted to larger values of $\mu$. On the other hand, in Case III with a $\mu$-dependent $T_0$,
a net rise of $\Phi$ occurs in correspondence of the chiral crossover. It would be very interesting if by means of the
DOS method one could compute the expectation value of the Polyakov loop, as well as the chiral and the Polyakov loop
susceptibilities, in the low temperature regime as a function of $\mu$. This lattice calculation might improve the
understanding of the new low temperature/large chemical potential state of matter claimed in~\cite{Fodor:2007vv}, and
at the same time it would allow to distinguish between the two PNJL scenarios discussed in this paper.

\section{Conclusions}
In this paper we have investigated on the landscape of the possible phases of the neutral two flavor PNJL model. We
have considered the logarithmic effective potential of the Polyakov loop ${\cal
U}$~\cite{Fukushima:2003fw,Roessner:2006xn}, see Eq.~\eqref{eq:Poly}, and a non local interaction in the quark sector,
see Eqs.~\eqref{eq:1}-\eqref{eq:f1}. Our main results are summarized in Figs.~\ref{fig:phased}
and~\ref{fig:phasedCOMP}. Fig.~\ref{fig:phased} corresponds to a fixed value of $\bar{T}_0$ in the Polyakov loop
effective potential. In this case the phase diagram is qualitatively similar to that obtained in the large $N_c$
approximation of QCD~\cite{McLerran:2007qj}.

In particular, at high chemical potential and low temperature we find a phase in which the main contribution to the
thermal quark population is given by $Z(3)$ neutral states, that is three quark states made of one red quark, one green
quark and one blue quark. This characteristic resembles the quarkyonic phase of Ref.~\cite{McLerran:2007qj}. The
quarkyonic-like structure of the ground state of the PNJL model has been already noticed in
Ref.~\cite{Fukushima:2008wg} in non neutral and three flavor version of the model. Moreover the $Z(3)$ transition line
has been already studied in Ref.~\cite{Sasaki:2006ww} with a different effective potential for the Polyakov loop and in
a non neutral state. The results of Ref.~\cite{Sasaki:2006ww} are qualitatively similar to ours. Therefore we suggest
that the quarkyonic-like state of matter is a feature of the PNJL model, independently to the number of flavors and to
the difference of the chemical potentials between quarks, as far as a $\mu$ dependence of the coefficients of ${\cal
U}$ is not considered.

In Fig.~\ref{fig:phasedCOMP} we show the phase diagram of the model when a $\mu$ dependence of the coefficients of the
effective potential of the Polyakov loop is introduced. We have used the analytic form suggested in
Ref.~\cite{Schaefer:2007pw}. The main results are the lowering of the $Z(3)$ transition line of Fig.~\ref{fig:phased},
and the shrinking of the quarkyonic-like phase window of Fig.~\ref{fig:phased}. We have used the form of
$\bar{T}_0(\mu)$ of Ref.~\cite{Schaefer:2007pw}. We believe that the result is rather robust as it does not follow from
such a detailed form but only from the lower scale of deconfinement when mu  increases.

We have not considered in this work for simplicity the possibility of color superconductivity at high
$\mu$~\cite{Rapp:1997zu,Alford:1998mk}. At a first sight it could seem that the results found with parameterizations I
and II, i.e. a quarkyonic-like phase at high chemical potential and low temperature, exclude the possibility of a
superconductive gap in the spectrum. This reasoning could be supported by the observation that the quarkyonic-like
phase is similar to a confined phase, differing from the latter only for a non zero value of the quark density. Such a
conclusion is not necessarily true. As a matter of fact, even if not noticed explicitly in Ref.~\cite{Roessner:2006xn}
for the two flavor and in~\cite{Ciminale:2007ei,Abuki:2008ht} for the three flavor models where the color
superconductivity has been kept into account, in the quarkyonic-like region (high $\mu$ and small $T$) the minimization
of the thermodynamic potential leads to a phase where quarks have a color superconductive gap in the spectrum. It is
the 2SC gap~\cite{Rapp:1997zu} in the two flavor case, and the CFL gap~\cite{Alford:1998mk} in the three flavor case.
Therefore the realization of a color superconductive phase in the PNJL models at high $\mu$ and small $T$ is not
forbidden in principle, even if the ground state has a quarkyonic structure.

An interesting investigation is the computation of the spectra of the mesonic and baryonic thermal excitations in the
quarkyonic-like phase of the PNJL model, and compare them with those obtained in a different
model~\cite{Glozman:2008kn} that mimics QCD in the large $N_c$ approximation. We are now working on this topic and the
results will be the object of a forthcoming paper.

\begin{acknowledgments}
We aknowledge K. Fukushima, M. Hamada, M. Huang, O. Kiriyama, T. Kunihiro, V. A. Miransky, C. Sasaki, A. Schmitt, I.
Shovkovy, W. Weise for discussions made during the Workshop ``New Frontiers in QCD 2008''. Moreover we have benefited a
discussion with K. Redlich during the aforementioned Workshop which has stimulated the main part of the present work.
We finally thank P. Cea and L. Cosmai for enlightening discussions, and D. Blaschke for useful correspondence.
\end{acknowledgments}



\begin{thebibliography}{99}

\bibitem{Aoki:2006br}
  Y.~Aoki, Z.~Fodor, S.~D.~Katz and K.~K.~Szabo,
  Phys.\ Lett.\  B {\bf 643}, 46 (2006)
  [arXiv:hep-lat/0609068].
\bibitem{Schmidt:2006us}
  C.~Schmidt,
  PoS {\bf LAT2006}, 021 (2006)
  [arXiv:hep-lat/0610116].
\bibitem{Philipsen:2005mj}
  O.~Philipsen,
  PoS {\bf LAT2005}, 016 (2006)
  [PoS {\bf JHW2005}, 012 (2006)]
  [arXiv:hep-lat/0510077].
\bibitem{Heller:2006ub}
  U.~M.~Heller,
  PoS {\bf LAT2006}, 011 (2006)
  [arXiv:hep-lat/0610114].

\bibitem{Ejiri:2004yw}
  S.~Ejiri,
  Phys.\ Rev.\  D {\bf 69}, 094506 (2004)
  [arXiv:hep-lat/0401012].
\bibitem{Splittorff:2006vj}
  K.~Splittorff,
  PoS {\bf LAT2006}, 023 (2006)
  [arXiv:hep-lat/0610072].
\bibitem{Splittorff:2006fu}
  K.~Splittorff and J.~J.~M.~Verbaarschot,
  Phys.\ Rev.\ Lett.\  {\bf 98}, 031601 (2007)
  [arXiv:hep-lat/0609076].


\bibitem{Allton:2003vx}
  C.~R.~Allton, S.~Ejiri, S.~J.~Hands, O.~Kaczmarek, F.~Karsch, E.~Laermann and C.~Schmidt,
  Phys.\ Rev.\  D {\bf 68}, 014507 (2003)
  [arXiv:hep-lat/0305007].

\bibitem{Allton:2002zi}
  C.~R.~Allton {\it et al.},
  Phys.\ Rev.\  D {\bf 66}, 074507 (2002)
  [arXiv:hep-lat/0204010].

\bibitem{Allton:2005gk}
  C.~R.~Allton {\it et al.},
  Phys.\ Rev.\  D {\bf 71}, 054508 (2005)
  [arXiv:hep-lat/0501030].

\bibitem{Fodor:2002km}
  Z.~Fodor, S.~D.~Katz and K.~K.~Szabo,
  Phys.\ Lett.\  B {\bf 568}, 73 (2003)
  [arXiv:hep-lat/0208078].



\bibitem{Fodor:2001pe}
  Z.~Fodor and S.~D.~Katz,
  JHEP {\bf 0203}, 014 (2002)
  [arXiv:hep-lat/0106002].


\bibitem{Fodor:2007vv}
  Z.~Fodor, S.~D.~Katz and C.~Schmidt,
  JHEP {\bf 0703}, 121 (2007)
  [arXiv:hep-lat/0701022].

\bibitem{Laermann:2003cv}
  E.~Laermann and O.~Philipsen,
  Ann.\ Rev.\ Nucl.\ Part.\ Sci.\  {\bf 53}, 163 (2003)
  [arXiv:hep-ph/0303042].
\bibitem{de Forcrand:2003hx}
  P.~de Forcrand and O.~Philipsen,
  Nucl.\ Phys.\  B {\bf 673}, 170 (2003)
  [arXiv:hep-lat/0307020].
\bibitem{D'Elia:2007ke}
  M.~D'Elia, F.~Di Renzo and M.~P.~Lombardo,
  Phys.\ Rev.\  D {\bf 76}, 114509 (2007)
  [arXiv:0705.3814 [hep-lat]].
\bibitem{D'Elia:2004at}
  M.~D'Elia and M.~P.~Lombardo,
  Phys.\ Rev.\  D {\bf 70}, 074509 (2004)
  [arXiv:hep-lat/0406012].
\bibitem{D'Elia:2002gd}
  M.~D'Elia and M.~P.~Lombardo,
  Phys.\ Rev.\  D {\bf 67}, 014505 (2003)
  [arXiv:hep-lat/0209146].


\bibitem{Nambu:1961tp}
  Y.~Nambu and G.~Jona-Lasinio,
  Phys.\ Rev.\  {\bf 122}, 345 (1961); 
  Phys.\ Rev.\  {\bf 124}, 246 (1961).


\bibitem{revNJL}
  S.~P.~Klevansky,
  Rev.\ Mod.\ Phys.\  {\bf 64}, 649 (1992);
  T.~Hatsuda and T.~Kunihiro,
  Phys.\ Rept.\  {\bf 247}, 221 (1994)
  [arXiv:hep-ph/9401310];
  M.~Buballa,
  Phys.\ Rept.\  {\bf 407}, 205 (2005)
  [arXiv:hep-ph/0402234].


\bibitem{Polyakovetal}
  A.~M.~Polyakov,
  Phys.\ Lett.\  B {\bf 72}, 477 (1978);
L.~Susskind,
  Phys.\ Rev.\  D {\bf 20}, 2610 (1979);
  B.~Svetitsky and L.~G.~Yaffe,
  Nucl.\ Phys.\  B {\bf 210}, 423 (1982);
  B.~Svetitsky,
  Phys.\ Rept.\  {\bf 132}, 1 (1986).



\bibitem{Meisinger:1995ih}
  P.~N.~Meisinger and M.~C.~Ogilvie,
  Phys.\ Lett.\  B {\bf 379}, 163 (1996)
  [arXiv:hep-lat/9512011].

\bibitem{Fukushima:2003fw}
  K.~Fukushima,
  Phys.\ Lett.\  B {\bf 591}, 277 (2004)
  [arXiv:hep-ph/0310121].

\bibitem{Ratti:2005jh}
  C.~Ratti, M.~A.~Thaler and W.~Weise,
  Phys.\ Rev.\  D {\bf 73}, 014019 (2006)
  [arXiv:hep-ph/0506234].
\bibitem{Roessner:2006xn}
  S.~Roessner, C.~Ratti and W.~Weise,
  Phys.\ Rev.\  D {\bf 75}, 034007 (2007)
  [arXiv:hep-ph/0609281].

\bibitem{Ghosh:2007wy}
  S.~K.~Ghosh, T.~K.~Mukherjee, M.~G.~Mustafa and R.~Ray,
  arXiv:0710.2790 [hep-ph];
  S.~K.~Ghosh, T.~K.~Mukherjee, M.~G.~Mustafa and R.~Ray,
  Phys.\ Rev.\  D {\bf 73}, 114007 (2006)
  [arXiv:hep-ph/0603050].

\bibitem{Kashiwa:2007hw}
  K.~Kashiwa, H.~Kouno, M.~Matsuzaki and M.~Yahiro,
  arXiv:0710.2180 [hep-ph].


\bibitem{Schaefer:2007pw}
  B.~J.~Schaefer, J.~M.~Pawlowski and J.~Wambach,
  Phys.\ Rev.\  D {\bf 76}, 074023 (2007)
  [arXiv:0704.3234 [hep-ph]].

\bibitem{Ratti:2007jf}
  C.~Ratti, S.~Roessner and W.~Weise,
  Phys.\ Lett.\  B {\bf 649}, 57 (2007)
  [arXiv:hep-ph/0701091].

\bibitem{Sasaki:2006ww}
  C.~Sasaki, B.~Friman and K.~Redlich,
  Phys.\ Rev.\  D {\bf 75}, 074013 (2007)
  [arXiv:hep-ph/0611147];
  C.~Sasaki, B.~Friman and K.~Redlich,
  Phys.\ Rev.\  D {\bf 75}, 054026 (2007)
  [arXiv:hep-ph/0611143].

\bibitem{Megias:2006bn}
  E.~Megias, E.~Ruiz Arriola and L.~L.~Salcedo,
  Phys.\ Rev.\  D {\bf 74}, 114014 (2006)
  [arXiv:hep-ph/0607338];
  E.~Megias, E.~Ruiz Arriola and L.~L.~Salcedo,
  Phys.\ Rev.\  D {\bf 74}, 065005 (2006)
  [arXiv:hep-ph/0412308].



\bibitem{Zhang:2006gu}
  Z.~Zhang and Y.~X.~Liu,
  Phys.\ Rev.\  C {\bf 75}, 064910 (2007)
  [arXiv:hep-ph/0610221].

\bibitem{Fukushima:2008wg}
  K.~Fukushima,
  arXiv:0803.3318 [hep-ph].

\bibitem{Sakai:2008py}
  Y.~Sakai, K.~Kashiwa, H.~Kouno and M.~Yahiro,
  Phys.\ Rev.\  D {\bf 77}, 051901 (2008)
  [arXiv:0801.0034 [hep-ph]].

\bibitem{Sakai:2008um}
  Y.~Sakai, K.~Kashiwa, H.~Kouno and M.~Yahiro,
  arXiv:0803.1902 [hep-ph].

\bibitem{Kashiwa:2008ga}
  K.~Kashiwa, Y.~Sakai, H.~Kouno, M.~Matsuzaki and M.~Yahiro,
  arXiv:0804.3557 [hep-ph].

\bibitem{Ciminale:2007ei}
  M.~Ciminale, G.~Nardulli, M.~Ruggieri and R.~Gatto,
  Phys.\ Lett.\  B {\bf 657}, 64 (2007)
  [arXiv:0706.4215 [hep-ph]].

\bibitem{Fu:2007xc}
  W.~j.~Fu, Z.~Zhang and Y.~x.~Liu,
  Phys.\ Rev.\  D {\bf 77}, 014006 (2008)
  [arXiv:0711.0154 [hep-ph]].

\bibitem{Ciminale:2007sr}
  M.~Ciminale, R.~Gatto, N.~D.~Ippolito, G.~Nardulli and M.~Ruggieri,
  arXiv:0711.3397 [hep-ph].

\bibitem{Hansen:2006ee}
  H.~Hansen, W.~M.~Alberico, A.~Beraudo, A.~Molinari, M.~Nardi and C.~Ratti,
  Phys.\ Rev.\  D {\bf 75}, 065004 (2007)
  [arXiv:hep-ph/0609116].

\bibitem{Abuki:2008tx}
  H.~Abuki, M.~Ciminale, R.~Gatto, N.~D.~Ippolito, G.~Nardulli and M.~Ruggieri,
  arXiv:0801.4254 [hep-ph].

\bibitem{Abuki:2008ht}
  H.~Abuki, M.~Ciminale, R.~Gatto, G.~Nardulli and M.~Ruggieri,
  arXiv:0802.2396 [hep-ph].

\bibitem{Contrera:2007wu}
  G.~A.~Contrera, D.~Gomez Dumm and N.~N.~Scoccola,
  Phys.\ Lett.\  B {\bf 661}, 113 (2008)
  [arXiv:0711.0139 [hep-ph]].

\bibitem{Blaschke:2007np}
  D.~Blaschke, M.~Buballa, A.~E.~Radzhabov and M.~K.~Volkov,
  arXiv:0705.0384 [hep-ph].

\bibitem{McLerran:2007qj}
  L.~McLerran and R.~D.~Pisarski,
  Nucl.\ Phys.\  A {\bf 796}, 83 (2007)
  [arXiv:0706.2191 [hep-ph]].
\bibitem{Hidaka:2008yy}
  Y.~Hidaka, L.~D.~McLerran and R.~D.~Pisarski,
  arXiv:0803.0279 [hep-ph].

\bibitem{Glozman:2007tv}
  L.~Y.~Glozman and R.~F.~Wagenbrunn,
  Phys.\ Rev.\  D {\bf 77}, 054027 (2008)
  [arXiv:0709.3080 [hep-ph]].
\bibitem{Glozman:2008kn}
  L.~Y.~Glozman,
  arXiv:0803.1636 [hep-ph].

\bibitem{Schmidt:1994di}
  S.~M.~Schmidt, D.~Blaschke and Yu.~L.~Kalinovsky,
  Phys.\ Rev.\  C {\bf 50}, 435 (1994).
\bibitem{Bowler:1994ir}
  R.~D.~Bowler and M.~C.~Birse,
  Nucl.\ Phys.\  A {\bf 582}, 655 (1995)
  [arXiv:hep-ph/9407336].
\bibitem{Blaschke:2000gd}
  D.~Blaschke, G.~Burau, Yu.~L.~Kalinovsky, P.~Maris and P.~C.~Tandy,
  Int.\ J.\ Mod.\ Phys.\  A {\bf 16}, 2267 (2001)
  [arXiv:nucl-th/0002024].
\bibitem{GomezDumm:2005hy}
  D.~Gomez Dumm, D.~B.~Blaschke, A.~G.~Grunfeld and N.~N.~Scoccola,
  Phys.\ Rev.\  D {\bf 73}, 114019 (2006)
  [arXiv:hep-ph/0512218].
\bibitem{Aguilera:2006cj}
  D.~N.~Aguilera, D.~Blaschke, H.~Grigorian and N.~N.~Scoccola,
  Phys.\ Rev.\  D {\bf 74}, 114005 (2006)
  [arXiv:hep-ph/0604196].

\bibitem{Grigorian:2006qe}
  H.~Grigorian,
  Phys.\ Part.\ Nucl.\ Lett.\  {\bf 4}, 223 (2007)
  [arXiv:hep-ph/0602238].

\bibitem{Ebert:2005wr}
  D.~Ebert and K.~G.~Klimenko,
  Eur.\ Phys.\ J.\  C {\bf 46}, 771 (2006)
  [arXiv:hep-ph/0510222].




\bibitem{Ebert:2000pb}
  D.~Ebert, K.~G.~Klimenko and H.~Toki,
  Phys.\ Rev.\  D {\bf 64}, 014038 (2001)
  [arXiv:hep-ph/0011273];
  D.~Ebert, V.~V.~Khudyakov, V.~C.~Zhukovsky and K.~G.~Klimenko,
  Phys.\ Rev.\  D {\bf 65}, 054024 (2002)
  [arXiv:hep-ph/0106110].

\bibitem{Ebert:2008tp}
  D.~Ebert, K.~G.~Klimenko, A.~V.~Tyukov and V.~C.~Zhukovsky,
  arXiv:0804.0765 [hep-ph].

\bibitem{Rapp:1997zu}
  R.~Rapp, T.~Schafer, E.~V.~Shuryak and M.~Velkovsky,
  Phys.\ Rev.\ Lett.\  {\bf 81}, 53 (1998)
  [arXiv:hep-ph/9711396];
  M.~G.~Alford, K.~Rajagopal and F.~Wilczek,
  Phys.\ Lett.\  B {\bf 422}, 247 (1998)
  [arXiv:hep-ph/9711395].
\bibitem{Alford:1998mk}
  M.~G.~Alford, K.~Rajagopal and F.~Wilczek,
  Nucl.\ Phys.\  B {\bf 537}, 443 (1999)
  [arXiv:hep-ph/9804403].








\end{thebibliography}
\end{document}